\documentclass[a4paper,fleqn]{cas-sc}

\usepackage[authoryear,longnamesfirst]{natbib}
\usepackage{subcaption}
\usepackage{pdflscape}
\usepackage{caption}
\usepackage{threeparttable}
\usepackage{booktabs}
\usepackage{hyperref}
\usepackage{xcolor}
\hypersetup{
    colorlinks=true,
    citecolor=blue,
    linkcolor=red,
    urlcolor=blue
}
\usepackage{lineno}

\def\tsc#1{\csdef{#1}{\textsc{\lowercase{#1}}\xspace}}
\tsc{WGM}
\tsc{QE}
\tsc{EP}
\tsc{PMS}
\tsc{BEC}
\tsc{DE}

\begin{document}

\shorttitle{Analyzing selected cryptocurrencies spillover effects on global financial indices}
\shortauthors{Rehman et~al.}

\title [mode = title]{Analyzing selected cryptocurrencies spillover effects on global financial indices: Comparing risk measures using conventional and $e$GARCH-EVT-Copula approaches}                      

\author[1]{Shafique Ur Rehman}
\affiliation[1]{organization={School of Economics and Management},
                addressline={University of Chinese Academy of Sciences}, 
                city={Beijing},
                country={China}}
\ead{shafique.u.r1000@mails.ucas.ac.cn}
\credit{Conceptualization, Methodology, Data curation, Software, Writing-Original draft preparation}

\author[2]{Touqeer Ahmad}[bioid=1]
\affiliation[2]{organization={CREST, ENSAI},
                city={Bruz},
                country={France}}
\ead{touqeer.ahmad@ensai.fr}
\credit{Writing-Original draft, Validation, Investigation, Reviewing, and Editing}

\author[1]{Wu Dash Desheng}[auid=000]
\cormark[1]
\ead{dwu@ucas.ac.cn}
\credit{Supervision, Review, and Editing}
\cortext[cor1]{Corresponding author}

\author[3]{Amirhossein Karamoozian}
\affiliation[3]{organization={International College},
                addressline={University of Chinese Academy of Sciences}, 
                city={Beijing},
                country={China}}
\ead{amir.hossein.karamoozian@ucas.ac.cn}
\credit{Visualization, Reviewing, and Editing}

\begin{abstract}
This study examines the interdependence between cryptocurrencies and international financial indices, such as MSCI World and MSCI Emerging Markets. We compute the value at risk (VaR), expected shortfall (ES), and range value at risk (RVaR) and investigate the dynamics of risk spillover. We employ a hybrid approach to derive these risk measures that integrates GARCH models, extreme value models, and copula functions. This framework is applied using a bivariate portfolio approach involving data from cryptocurrencies (Bitcoin and Litecoin) and traditional financial indices. To estimate the above risks of these portfolio structures, we employ symmetric and asymmetric GARCH and both tail flexible EVT models as marginal to model the marginal distribution of each return series and apply different copula functions to connect the pairs of marginal distributions into a multivariate distribution. The empirical findings indicate that the \texttt{eGARCH} EVT-based copula model adeptly captures intricate dependencies, surpassing conventional methodologies like Historical simulations (HS) and $t$ distributed parametric in VaR estimation. At the same time, the HS method proves superior for ES, and the $t$ distributed parametric method outperforms RVaR. Eventually, the Diebold-Yilmaz approach will be applied to compute risk spillovers between four sets of asset sequences. This phenomenon implies that cryptocurrencies reveal substantial spillover effects among themselves but minimal impact on other assets. From this inference, it can be concluded that cryptocurrencies propose diversification benefits and do not provide hedging advantages within an investor's portfolio. 
Furthermore, our results underline RVaR superiority over ES regarding regulatory arbitrage and model misspecification. The conclusions of this study will benefit investors and financial market professionals who aspire to comprehend digital currencies as a novel asset class and attain perspicuity in regulatory arbitrage.
\end{abstract}


\maketitle


\section{Introduction}

Intermarkets play a crucial role in the functioning of financial markets, particularly at the international level. The interrelationship between commodity and equity markets is of significant importance in risk spillover analysis. This is especially pertinent for portfolio executives who prioritize global indices and are vigilant about financial crises. Understanding these spillovers allows for more informed decision-making and effective risk management in diversified portfolios. In recent years, cryptocurrencies have garnered significant attention from researchers and inventors, among other financial investments \citep{gil2020cryptocurrencies, ugolini2023connectedness}.  Cryptocurrencies are associated with a higher market risk due to their greater volatility compared to traditional financial assets \citep{ghosh2023return, nguyen2023volatility, lazar2024estimation}.  For example, investors in these markets experienced severe losses during the 2018 cryptocurrency market downturn. The values of major cryptocurrencies plummeted dramatically from their peak in January 2018 to their lows in December 2018.\footnote{\href{https://coinmarketcap.com/}{https://coinmarketcap.com/}}. 
\par Over the past few years, cryptocurrencies have become increasingly popular. For instance, 20,000 cryptocurrencies were traded in August 2022, compared to 600 in January 2016. Consequently, the cryptocurrency market boasts a capitalization of one trillion USD \citep{baer2023taxing}. Traders and investors, seeking to diversify their assets and manage financial risks, often turn to cryptocurrencies as a hedge against market uncertainties \citep{jiang2021revisiting, riahi2024investing}. The interconnection between the stock market and cryptocurrencies warrants further discussion in the existing literature, particularly regarding the integration of cryptocurrencies with conventional financial assets \citep{ferko2023trades}. The global financial crisis in 2008 and the COVID-19 pandemic in 2019 have significantly disrupted the global economy, posing challenges for investors and traders in developing risk management portfolios that include cryptocurrencies. Given these events, analyzing market volatility and spillover effects is crucial for financial professionals and policymakers. Quantifying market risk and spillovers between financial assets remains a significant challenge for mathematicians, academicians, financial executives, and policymakers. In the context of financial markets, risk, and spillovers refer to the phenomenon where a loss in one asset or a set of assets, or a crisis in one country, escalates the risk associated with other countries and assets \citep{gkillas2019integration}.
\par In contemporary financial practice and academic discourse on risk forecasting, Value at Risk (VaR) and Expected Shortfall (ES) have emerged as pivotal methodologies for quantifying market risk \citep{basel2013fundamental}.
Despite widespread adoption, both VaR and ES exhibit specific deficiencies. VaR, for instance, lacks convexity and may undermine the benefits of diversification \citep{artzner1999coherent}. Additionally, VaR neglects quantiles below a specified threshold of interest. In contrast, ES addresses these deficiencies by accounting for all quantiles below a given significance level, thus demonstrating convexity. However, ES overlooks two crucial statistical properties inherent to VaR: robustness and elicitability. Robustness accounts for model uncertainty, while elicitability facilitates comparison among competing models. \citep{cont2010robustness} proposed the Range Value at Risk (RVaR) as a robust adaptation of ES, presenting enhanced suitability for practical applications while preserving the fundamental nature of ES relative to VaR and ES alternatives. Despite being introduced over a decade ago, the literature on forecasting approaches for RVaR remains nascent. Recently \citep{fissler2021elicitability, muller2022comparison} have pioneered efforts to explore combined elicitability and propose a loss function for comparing alternative forecasting strategies for RVaR. In the current study, we aim to forecast VaR, ES, and RVaR using diverse methodologies, including non-parametric, parametric, and Monte Carlo simulation techniques. Subsequently, we evaluate their performances by examining average realized loss values. Furthermore, it explores model misspecification and legal robustness to determine models that accurately apprehend the true underlying risk process, presenting benefits for regulatory arbitrage as documented by \citep{kellner2016quantifying}.
\par To address the flaws of conventional methods, we propose hybridizing diverse mathematical models to achieve greater accuracy in risk forecasts. Models based on Extreme Value Theory (EVT) have become seen to be favorite candidates for scrutinizing the sequences and appropriately exploring the risk spillover effects between cryptocurrencies and traditional financial markets. For instance, prior research predominantly focused on Bitcoin as a representative of cryptocurrencies, ignoring others like Ethereum and Litecoin, which also retain significant market existence and affirm thorough assessment. Therefore, the passionate alliance of cryptocurrencies with global financial indices is helpful for financial managers, critics, policymakers, and other market participants. With regard to the existing literature, it is imperative to conduct a comprehensive analysis of the behavior of significant financial assets and associations concerning the severity of the impact of risk spillovers. The present study addresses these issues by implementing novel methods.
\par For this study, Bitcoin, Litecoin, Morgan Stanley Capital International World Index (MSCI World), and Morgan Stanley Capital International Emerging Index (MSCI Emerging) are selected to form a portfolio comprising duos (Bitcoin-MSCI World, Bitcoin-MSCI Emerging, Litecoin-MSCI World, Litecoin-MSCI Emerging) for an in-depth inspection of their behavior. To assess the VaR, ES, and RVaR of these portfolios, we applied the Generalised Auto-regressive Conditional Heteroscedastic (GARCH) along with the EVT-Copula approach. The framework follows the following steps. (1) Compare different symmetric (sGARCH) and asymmetric (eGARCH, gjrGARCH) GARCH models. The best-fitted model is identified based on the Akaike Information Criterion (AIC), Bayesian Information Criterion (BIC), Root Mean Squared Error (RMSE), and Mean Absolute Error (MAE) values, adequately capturing the volatility of each data series. (2) Compare different extreme value heavy-tailed probability distributions such as Bulk-and-Tails (BATs), Generalized Pareto-Normal-Generalized Pareto (GNG), and the Cauchy distribution. The optimal distribution is determined for further analysis. (3) Use the specified distribution as marginal to construct a distinct copula function. (4) Apply the MCS procedure to acquire the VaR, ES, and RVaR estimates, which are evaluated based on their average realized loss values. (5)  Compare the risk measures (VaR, ES, and RVaR) obtained from the GARCH-EVT-Copula framework with those obtained using conventional estimation methods (non-parametric, parametric). As far as we are aware, no such study has been found in the literature that utilized novel lower and upper tails flexible distribution (i.e., BAT model) in conjunction with different classes of copulas. The proposed approach exhaustively interprets the VaR, ES, and RVaR based on the GARCH-EVT-Copula framework for cryptocurrencies (Bitcoin, Litecoin) and world financial Indices (MSCI\_W, MSCI\_EM).
\par Moreover, this study investigates the spillover effects of the cryptocurrencies to furnish better investment opportunities to simplify the risk management in cryptocurrencies, MSCI\_W, and MSCI\_EM. By using the proposed method, we achieve new and more effective results.The findings suggest that integrating EVT with the structural interdependence among the chosen time series enhances the effectiveness of risk assessment among the specified assets concerning their distributional attributes. In addition, the risk spillover analysis reveals that cryptocurrencies are highly correlated with each other but exhibit insignificant correlations with other assets. This phenomenon supports the consideration of cryptocurrencies in portfolio optimization. Traders and investors can manage higher or lower levels of volatility more effectively by incorporating both digital and traditional assets in their portfolios.
\par  The significant contributions to the literature of the current study are as follows: first, our investigations evaluate the risk spillover impacts among digital and traditional assets. Secondly, we conduct a comparative analysis between various GARCH models (sGARCH, eGARCH, gjrGARCH) and extreme value heavy-tailed probability distributions (BATs, GNG, Cauchy), followed by the integration of the best-specified models into a hybrid approach based on copulas (Frank, Gumbel, Joe, and Student's t), resulting in the \texttt{eGARCH} EVT-Copula framework. This demonstrates a novel methodology. This study also focuses on the use of both tails flexible model (i.e., BATs model), which is developed to capture both tail behaviors without selection of any threshold near the tail. Thirdly, we compute each portfolio's risk measures, employing conventional and hybridized methods, confining VaR, ES, and RVaR. Fourthly, we evaluate the performance of various models by estimating their scoring functions, comparing VaR, ES, and RVaR regarding legal robustness and model misspecification, and propose novel practical insights for industrial and empirical investigations.
\par Rest of the paper is arranged as follows. Sections \ref{Lit_review} provide an extensive literature review, and section \ref{Mehodology} describes the methodological framework, including the GARCH-EVT based Copula model, risk spillover estimations formulas, copula modeling, and the procedure employed for computation of VaR, ES, and RVaR, their scoring functions, and model misspecification and legal robustness. Section \ref{applications}  deals with the empirical application and discusses the results. Section \ref{consclusion} concludes the study by recapitulating the principal findings and furnishing policy implications and forthcoming proposals.
\section{Literature Review}\label{Lit_review}

Numerous studies in the literature have investigated the transmission and spillover effects of cryptocurrency prices and risks, as well as their associations with various economic and financial assets across different markets. For instance, research has been conducted on the relationship between cryptocurrencies and traditional financial assets \citep{mizerka2020role,parfenov2022efficiency, kyriazis2023can,bhanja2023aggregate, mensi2023quantile, huang2024relationship}. Additionally, some studies have examined return and volatility spillover mechanisms to understand the impact of crises and disturbances on other markets \citep{uzonwanne2021volatility, yaya2022persistence, hanif2023volatility, liu2024unveiling, kyriazis2024quantifying}. Exploring EVT and employing copula functions in financial analysis represents another line of inquiry to evaluate risk estimation and interdependence among different asset classes \citep{hussain2020dependence, jeribi2021portfolio,  zhao2023extreme, hanif2023dependence, karimi2023analyzing}.  A separate study has also been classified for comprehensive examination in order to garner comprehensive insights into the practical applications and properties of VaR, ES, and RVaR across diverse categories \citep{cont2010robustness, gneiting2011making, bellini2015elicitable, muller2022comparison, biswas2023nonparametric, muller2023comparison}. There are the subsequent study classes in which we prolonged to illustrate each relevant concept.

\subsection{\textnormal{\textit{Cryptocurrencies and Global Financial Indices}}\label{2.1}}
\citet{mizerka2020role} investigate the correlation between stock market assets and attributes of Bitcoin's user graph, focusing on the substantial number of Bitcoin users encountered in transactions with the highest amounts. Their findings verified a significant relationship, if present, and highlighted its notable impact on the behavior of Bitcoin returns for its users. \citet{kyriazis2023can} examine the non-linear causal relationship across various quantile levels through which cryptocurrencies provide a shelter for global financial indices. Their findings suggested that cryptocurrencies have a parallel capability to long-standing traditional financial assets, which provide the hedging mechanism by considering the extreme fluctuations of investors' portfolios.
\citet{bhanja2023aggregate} explore the portfolio diversification across various asset classes (cryptocurrency, equity, and precious metals) employing the frequency-based spillover transmission mechanism. They encountered that overall spillover effects suggest that Bitcoin is a doable diversification opportunity distinct from other asset classes, while the breakdown of total spillover reveals asymmetric inter-dependencies among these markets. \citet{kyriazis2024quantifying} investigates the dynamic inter-connectedness between precious metals, industrial metals, oil, natural gas, and Bitcoin using the Quantile Vector Auto-regressive methodology. Their findings provide insights into how cryptocurrency and commodities markets can attract new investors and assist in mitigating severe economic and financial crises.

\subsection{\textnormal{\textit{Risk spillover analysis among cryptocurrencies and global financial indices}}\label{2.2}}
\citet{uzonwanne2021volatility} verified the existence of returns and volatility spillovers between Bitcoin and major stock markets and quantified risk spillover employing the innovative multivariate VARMA-AGARCH model. They determined both uni-directional and bi-directional volatility spillovers in certain markets, signifying that investors may shift between them to optimize returns and mitigate risk vulnerability. \citet{yaya2022persistence} examine the volatility spillovers from bitcoin to sliver and gold market performing the fractional persistence mechanism. They illustrate the bivariate spillover effect using the CCC-VARMA-GARCH approach and conclude that there is no spillover effect among bitcoin returns to gold (silver). \citet{liu2024unveiling} examine the connectedness between cryptocurrencies implied exchange rate discounts in BRICKS (Brazil, Russia, India, China, and South Africa) and the US financial market by employing a connectedness approach based on time time-varying parameters Auto-regression and DCC-GARCH. Their outcomes imply that cryptocurrencies' inferred exchange rate spillover effects are more robust than the total spillover effect and contribute most to market connectedness.

\subsection{\textnormal{\textit{Understanding the EVT based copula functions}}\label{EVTbasedcopula}}

Indeed, it is acknowledged that while the probability of occurrence of extreme circumstances is small, the potential consequences of extreme circumstances on financial markets can be enormous. Consequently, we desire to pay attention to the tail-related characteristics of financial assets. By overwhelming the issue of the distribution hypothesis, EVT presents a specified statistical approach to model the distribution's tails and authorizes the computation of extreme risk measurements, such as return level, value at risk, and expected Shortfall. 
\citet{hussain2020dependence} investigate the dependence structure between stock market indices and cryptocurrencies by employing the time-varying copula and EVT model. The results of their research offer valuable insights for investors and portfolio managers aspiring to improve their understanding of diversification across diverse assets.
\citet{jeribi2021portfolio} proposed the two phases of the methodology. First, they examine the affinity between cryptocurrencies, oil prices, and US stock indices. Secondly, they determine the finest portfolio hedging approach and employ the FIEGARCH-EVT-Copula and hedging ratios framework. The findings unveil a significant negative leverage effect in US indices and crude oil, alongside a positive asymmetric volatility apparent in cryptocurrency markets. 
\citet{zhao2023extreme} explore the risk spillover influence of oil prices on the Chinese stock markets by utilizing the GARCH-EVT-Copula-CoVaR model. They address the tail dependence, volatility clustering, and more promising extreme circumstances to investigate risk estimation more robustly. Their findings demonstrate diverse spillover effects across sectors, with a notable positive spillover from the crude oil market to other Chinese stock markets, highlighting the effectiveness of their proposed model in enhancing risk assessment.
\par The dependence layout and portfolio disbursement method are examined between the industrial portfolio metals (gold, platinum, palladium, aluminum, silver, copper, zinc, lead, and nickel) and agricultural commodities portfolio (wheat, corn, soybeans, coffee, sugar cane, sugar beets, cocoa, cotton, and lumber) by employing the vine copula based conditional value at risk model \citep{hanif2023dependence}. The outcomes imply symmetric dependence dynamics in the leading metal portfolio, asymmetric and symmetric characteristics in the agricultural commodities portfolio, and a small degree of riskiness in the metal commodities portfolio during financial crises. \citep{karimi2023analyzing} explore the dependency structure and evaluate the VaR and risk spillovers among the cryptocurrencies, gold, and oil prices by employing the Diebold-Yilmaz and GJR-GARCH-EVT model.
The results suggest that the GJR-GARCH-EVT model outperforms traditional methods for risk quantification, with minimal spillover estimates for oil and gold. This indicates that cryptocurrencies are effective hedges against these assets and appropriate diversifiers for investors' portfolios.

\subsection{\textnormal{\textit{Exploration of risk measures}}\label{2.3}}

In the literature, numerous studies have focused on the applications of VaR and ES despite their inherent limitations. Addressing these constraints, \citet{cont2010robustness} introduced two parametric Range Value at Risk measures, demonstrating notable robustness compared to VaR and ES. \citet{muller2022comparison}  forecasted leading cryptocurrencies employing GARCH models with varying distributions of innovation. They suggested that the primary determinant of assets' RVaR is the conditional standard deviation rather than the distribution of stochastic terms, and they found that non-normal distributions perform well, as assessed through scoring functions. \citet{biswas2023nonparametric}  explored the accuracy of multiple non-parametric estimators of RVaR across various scenarios employing Monte Carlo simulation methods. Their methodology investigated the impact of altering the order of $p$ and $q$ relative to $n$ on the effectiveness of RVaR estimators and also executed the backtesting activity. Their results imply that their simulation study and backtesting approaches demonstrate that the non-parametric estimator of RVaR, defined by the filtered historical estimator of ES, surpasses other estimators in most scenarios. \citet{muller2023comparison} assess the effectiveness of diverse models for predicting VaR, ES, and RVaR through both univariate and multivariate analyses. This evaluation process incorporates distinct asset classes, rolling window estimations, and significance levels. They computed the scoring functions for each measure, finding that the GARCH model with $t$ distribution and skewed error distribution outperformed other distributions in univariate analysis. Conversely, the Rvine and Cvine copulas demonstrated superior performance in multivariate analysis.

\par In sum, the literature review underlines the significance of our research and furnishes a more extended and concentrated analysis of risk spillover developments among cryptocurrencies and global financial indices. Moreover, the literature also assesses the consequences of risk measures, including their extreme aspects, and proposes worthwhile recommendations for investors, portfolio managers, and researchers.

\section{Methodology}\label{Mehodology}

This section explains the mathematical construction of the models used in this study. We first describe the different GARCH  (symmetric and asymmetric) models. Then the extreme value marginal distributions are discussed which later used to construct extreme value copula functions. Then, by employing the best-fitted models, we integrate the GARCH-EVT-based Copula approach, which would predict the VaR, ES, and RVaR of portfolios among cryptocurrencies and global financial indices. Eventually, we compute the risk spillovers among variables.

\subsection{\textnormal{\textit{Volatility models}}\label{Volatility_model}}
Volatility clustering is a common phenomenon in financial time series, where high-volatility events tend to cluster together. In practical applications, the choice between symmetric and asymmetric GARCH models depends on the data and the specific characteristics of the asset being analyzed. Symmetric GARCH models provide a basic framework, while asymmetric models like offer more nuanced insights by accounting for the differential impact of positive and negative shocks on volatility. By leveraging these models, researchers and practitioners can better manage financial risks and make more informed investment decisions. Some useful models considered for this study are discussed subsequently.

\subsubsection{\textnormal{\textit{\texttt{sGARCH(p,q)} model}}\label{3.1.1}}

The major drawback of the Auto-regressive Conditional Heteroskedasticity (\texttt{ARCH(p)}) model is the lack of a definitive strategy for specifying the order of the model. Specifying the order $p$
is crucial as it involves selecting the number of parameters for the model. To address this issue, \citet{bollerslev1986generalized} introduced the Generalized Auto-regressive Conditional Heteroskedasticity (\texttt{GARCH} or \texttt{sGARCH}) model, which mitigates the limitations of the \texttt{ARCH} model. The conditional variance equation of the \texttt{GARCH} model, similar to the \texttt{ARCH} model, is expressed as:
\begin{equation}\label{sGARCH}
\begin{split}
\texttt{sGARCH(p,q)}: h_t = \gamma_{0} + \sum_{i=1}^p \delta_{i} h_{t-i} + \sum_{j=1}^q \gamma_{j} \mu_{t-j}^2,
\end{split}
\end{equation}

where 
$\mu_{t-j}$
  represents the residual terms, and 
$h_t$
  is the conditional variance, which is predicted by its own lagged values and the squared residuals.

\subsubsection{\textnormal{\textit{\texttt{eGARCH(p,q)} model}}\label{3.1.2}}

To more comprehensively analyze the distinct impacts of positive and negative shocks within the time series, the \texttt{eGARCH} model has been employed as a novel approach. The \texttt{eGARCH} model was proposed by \citet{nelson1991conditional}, which specifies the conditional variance, and the equation is expressed as
\begin{equation}\label{GARCH}
    \texttt{eGARCH}:\log(h_t) = \alpha + \sum_{j=1}^q \gamma_{j} \left| \left( \frac{u_{t-j}}{\sqrt{h_{t-j}}} \right) \right| + \sum_{j=1}^w \phi_{j} \left( \frac{u_{t-j}}{\sqrt{h_{t-j}}} \right)
    + \sum_{i=1}^p \delta_i h_{t-i}
 \end{equation}
The positive values of $\gamma_{j}$ and $\phi_{j}$ imply that both positive and negative news have a parallel influence. Furthermore, the negative values of $\phi_{j}$ coupled with the positive values of $\gamma_{j}$ demonstrate that positive shocks exert a less significant impact on the time series than adverse shocks. The \texttt{eGARCH} model is valuable for assessing how shocks, based on their magnitude and sign, exploit conditional volatility.
\subsubsection{\textnormal{\textit{\texttt{gjrGARCH(p,q)} model}}\label{3.1.3}}
The \texttt{gjrGARCH} model is the simplest type of asymmetric \texttt{GARCH} and the first presented by \citet{nelson1991conditional}. The conditional variance equation is expressed as
\begin{equation}\label{gjrGARCH}
\texttt{gjrGARCH(p,q)}:h_t = \alpha_{0} +\sum_{i=1}^p\alpha_{i}\mu_{t-i}^2 + \sum_{i=1}^p\gamma_{i} S_{t-i} \mu_{t-i}^2+ \sum_{j=1}^q\beta_{i} h_{t-j}^2,
 \end{equation}
 where $S_{t-i}$ portrays the dummy variable, it becomes zero if $\mu_{t-i}$ is positive otherwise 1. In this procedure, the influences of positive shocks are equal to $\alpha_{i} \mu_{t-i}^2$, and the impact of adverse shocks is equal to $(\alpha_{i}+\gamma_{i})\mu_{t-i}^2$. Since bad news has a more significant effect on variance, this is projected to have a positive $\gamma_{i}$.
\subsection{\textnormal{\textit{Modeling extreme events in risk management}}\label{Riskmodel}}
In the domain of risk management, extreme events can lead to severe disturbances. Despite the decreasing likelihood of these occurrences, their potential costs and significant consequences demand meticulous assessment. Notable examples of such events include substantial market crashes, the failure of large institutions to meet their liabilities, financial market catastrophes, and hurricanes. Evaluating these occurrences and estimating extreme-value risk metrics is a crucial aspect of risk management. This study simultaneously integrates the outcomes of the \texttt{GARCH} model with Extreme Value Theory (EVT) models to address heteroscedasticity and the occurrence of extreme events.

Two widely employed EVT approaches are the Block Maxima (BM) method and the Peak Over Threshold (POT) method. The BM approach partitions the dataset into equal non-overlapping blocks and extracts the maximum and minimum values from each block. However, to address the limitations of the BM approach, the POT method provides a more detailed perspective. It focuses on observations that exceed a predefined threshold, thereby facilitating the modeling of the tail distribution, which has garnered significant attention in risk management in recent years \citep{karmakar2019intraday, chebbi2022revisiting}.  In both EVT approaches, most of the data is disregarded, and only the tails of the distributions are used for inference. Extreme events are characterized as values exceeding a threshold or as maximum (or minimum) values over distinct time blocks, inspired by the asymptotic theory for extremes of stochastic processes. To accurately estimate the extent of these extreme observations, it is necessary to consider the entire dataset rather than just the threshold or maximum (minimum) values. This study utilizes a newly proposed parametric model for the underlying probability distribution, which exhibits flexible behavior in both tails. 
By integrating the \texttt{GARCH} model, we can effectively account for heteroscedasticity and extreme events. The \texttt{GARCH} model captures time-varying volatility, while the EVT models focus on the tails of the distribution, providing a comprehensive framework for estimating extreme-value risk metrics in financial and risk management contexts.
\subsubsection{\textnormal{\textit{Extreme value model with flexible heavy tails}}\label{bat}}

In EVT literature, the Generalized Pareto distribution is a common choice to estimate the extreme upper quantiles of distributions for a small fraction of the largest observations. In this proposed study, we are interested in both tails of the distributions of the returns obtained through volatility models. For modeling both tails of return distributions, we rely on both tails flexible extreme value parametric model known as bulk-and-tails (BATs) distribution. This model is recently introduced in extreme value literature by \citet{stein2021parametric}. We construct the model by assuming $T_\nu(H_\theta(x))$ is a cumulative distribution (CDF) of BATs, where $T_\nu$ is the cdf of a student-t random variable with $\nu$ degree of freedom and $H_\theta$ is the monotone-increasing function having six parameters that restrain the upper and lower tails. Define $\Psi(x)=\log(1+e^{x})$, the CDF of BAT distribution is defined as
\begin{equation}\label{BATs1}
H_{\theta}(x) = \left\{1 + \kappa_{1} \Psi \left(\frac{x - \varphi_{1}}{\tau_{1}}\right)\right\}^{\frac{1}{\kappa_{1}}} - \left\{1 + \kappa_{0} \Psi \left(\frac{\varphi_{0} - x}{\tau_{0}}\right)\right\}^{\frac{1}{\kappa_{0}}},
 \end{equation}
where $(\varphi_{1},\tau_{1},\kappa_{1})$ are the location, scale, and shape parameters for the upper of the distribution while  $(\varphi_{0},\tau_{0},\kappa_{0})$ are same parameters for the lower tail of the distribution. The parameters $\kappa_{i}, i=0,1$ control the tails behavior of the distribution. For instance, positive values represent the heavy tail while negative values show the thin tail with bounded support.  Distributions, respectively, with restricted support to those tails. Let $(L, U)$ demonstrate the interior of the support. if $\kappa_{1}\geq0$ the upper limit is $U=\infty$, while if $\kappa_{1}<0$, then $U=\varphi_{1}+\tau_{1}\Psi^{-1}(-1/\kappa_{1})$. similarly $L=-\infty$ , if $\kappa_{0}\geq0$ and $L=\varphi_{0}-\tau_{0}\Psi^{-1}(-1/\kappa{0})$, if $\kappa_{0}<0$. If $\kappa_0,\kappa_1\rightarrow0$ in \eqref{BATs1}, we simply get
\begin{equation*}\label{BATs2}
H_\theta(x)=\exp\left(\frac{x-\varphi_{1}}{\tau_{1}}\right)-\exp\left(\frac{\varphi_{0}-x}{\tau_{0}}\right)
\end{equation*}
The probability density function is obtained by differentiating the CDF with respect to $x$ as
\begin{equation}\label{BATspdf}
\begin{aligned}
f_{\theta}(x)={}t_{\nu}(H_{\theta}(x))\Bigg[\frac{1}{\tau_1}\left\{1 + \kappa_{1} \Psi \left(\frac{x - \varphi_{1}}{\tau_{1}}\right)\right\}^{\frac{1}{\kappa_{1}}-1}\Psi' \left(\frac{x - \varphi_{1}}{\tau_{1}}\right)+ \frac{1}{\tau_0}\left\{1 + \kappa_{0} \Psi \left(\frac{\varphi_{0} - x}{\tau_{0}}\right)\right\}^{\frac{1}{\kappa_{0}}-1}
\Psi' \left(\frac{\varphi_{0} - x}{\tau_{0}}\right)\Bigg],
\end{aligned}
 \end{equation}
where $t_{\nu}$ is the density function for $T_{\nu}$ with $\nu$ degrees of freedom. For estimation purposes, the likelihood function is defined as 
\begin{equation}\label{BATs3}
\Pi_{i=1}^m f_{\theta}(x_i)
\end{equation}  
Defining the fitting and working strategy of BATs is obtainable through the Julia package BulkAndTails.jl, which is also available for the R language. The package incorporates the code for distributing seven standard parameters of BATs \citep{krock2022nonstationary}.
\subsubsection{\textnormal{\textit{Generalized Pareto-
Normal-Generalized Pareto (GNG)} model}\label{3.2.2}}

There is another model that is favorite for modeling the entire distribution of financial residual time series. This model mixes the two distributions, the normal is used for modeling the central part of residual distribution while tails are modeled through GPD distribution. \citet{macdonald2011flexible} proposed this model to jointly the distribution of data and this model has been widely used in stock market return modeling \citep{ghanbari2019coherent, karimi2023analyzing}.
To fit this model, a threshold is required to define the tails of the data distribution. However, the selection of an optimal threshold presents a challenge. This is a kind of disadvantage of this model. We take this model till an end and compare its performance with a novel model presented in section  \ref{bat}. The GNG model is defined by its CDF as
\begin{equation}\label{GNG}
F(x|\mu,\sigma,u,\alpha,\xi,\phi)= 
\begin{cases}
\frac{1-\phi}{1+\text{erf}\left(\frac{u-\mu}{\sigma\sqrt{2}}\right)}\left[1+\text{erf}\left(\frac{x-\mu}{\sigma\sqrt{2}}\right)\right]  \text{if } x<u [10pt]
(1-\phi)+\phi\left[1-\left(1+\xi\frac{x-u}{\alpha}\right)^\frac{1}{\xi}\right] \text{if } x\geq u, 
\end{cases}
\end{equation}
where $u$, $\alpha$, and $\xi$ express the location, scale, and shape parameters of the GPD and $\mu, \sigma$ denotes the location and scale parameters of normal distribution. At the same time, $\phi$ denotes the probability of independent exceedances over the threshold, and $\texttt{erf}$ defines the error function. The maximum likelihood procedure is employed to estimate the parameters of the model.

\subsubsection{\textnormal{\textit{Cauchy distribution}}\label{3.2.3}}
The Cauchy distribution, classified as stable, demonstrates heavy tails and claims similarities with the Pareto distribution \citep{VOS201875}. \textcolor{black}{The principal limitation inherent in the Cauchy distribution lies in its scarcity of finite moments, inducing it inadequate for accurately modeling financial series as it fails to retain finite $r^{th}$ moments \citep{patterson2003impact}.} The PDF of the Cauchy distribution is given as
\begin{equation}
    f(x;x_{0},\delta)=\frac{1}{\pi\delta\left[1+(\frac{x-x_{0}}{\delta})^2\right]}=\frac{1}{\pi}\left[\frac{\delta}{(x-x_{0})^2+\delta^2}\right]
\end{equation}
 and CDF is defined as
\begin{equation}
    F(x;x_{0},\delta)=\frac{1}{\pi}\textit{arctan}\left(\frac{x-x_{0}}{\delta}\right)+\frac{1}{2}.
\end{equation}
We compare BAT, GNG, and Cauchy models to obtain the most appropriate probability distribution for the marginal configuration of the copula function which we defined later for modeling dependence.

\subsection{\textnormal{\textit{Risk measures}}\label{3.3}}

In finance research, risk refers to the degree of uncertainty and the potential for financial loss associated with an investment decision. In this study, a variety of quantitative measures of risk are examined. For instance, the financial positions are represented as random variables based on their net present value. Let $\mathcal{G}:=L^1(\Omega,\mathcal{F}, \mathcal{P})$ is a space of all 
$\mathcal{F}$-measurable functions on 
$\Omega$ that are integrable with respect to the probability measure 
$\mathcal{P}$, which allows us to define the quantitive risk measure on a given space. The real number $X(\omega)>0$ is an "ex-post" gain, while $X(\omega)<0$ be "ex-ante", the certain losses. For all $X\in \mathcal{G}$, the CDF, expected value, and left quantile are defined $ F_X(x):= \mathbb{P}(X \leq x)\quad\forall x \in \mathbb{R}, \mathbb{E}[X]:= \int_{\Omega} X \, d\mathbb{P}\quad\forall X \in \mathcal{G}, $ and $ F_X^{-1}(\alpha) = \inf\{x \in \mathbb{R}: F_X(x) \geq \alpha\}\quad \forall\alpha\in(0,1)$,
respectively. The $\rho:\mathcal{G}\xrightarrow{}\mathbb{R}$ indicates that the risk measure 
$\rho$ maps each element of 
$\mathcal{G}$ (each integrable function representing a financial position) to a real number. In simpler terms, a risk measure 
$\rho$ assigns a real number to each financial position or portfolio $X\in \mathcal{G}$. This real number represents the risk of that position. The exact nature of this risk value depends on the specific properties and definition of the risk measure $\rho$. There are several desirable properties that a risk measure might possess, especially in the context of coherent risk measures, as introduced in \citet{artzner1999coherent}, some other theoretical aspects of risk measures are also discussed in \citet{delbaen2002coherent, frittelli2004dynamic}. This study considers the following definitions.\begin{enumerate}
\item  For $\alpha\in(0, 1)$ level of significance, the VaR is defined as
\begin{equation}\label{var11}
\text{VaR}^{\alpha}(X):= F^{-1}_{-X}(1-\alpha)\equiv -\text{inf}\{x\in\mathbb{R}:F^{-1}_{X}(x)\geq\alpha\}\equiv-F^{-1}_{X}(\alpha), \forall X\in \mathcal{G}
\end{equation}
\item  For the same level of significance, the ES is defined as
\begin{equation}\label{es}
    \text{ES}^{\alpha}(X):= \frac{1}{\alpha}\int^{\alpha}_{0} \text{VaR}^{u}(X)du, \forall X\in \mathcal{G}
\end{equation}
\end{enumerate}
The VaR\(^\alpha\) does not take into account the information below the \(\alpha\)-quantile, which is considered a significant drawback. This limitation means that VaR only measures the threshold value beyond which losses occur but does not provide insight into the potential severity of losses beyond this threshold. On the other hand, ES addresses this issue by considering the average of all potential losses that occur below the \(\alpha\)-quantile. ES not only provides a more comprehensive risk assessment but also adheres to the axioms of coherency as proposed by \citet{artzner1999coherent}. These axioms ensure that ES is a coherent risk measure, meaning it satisfies desirable properties such as subadditivity, which ensures that the risk of a diversified portfolio (comprising multiple assets) is less than or equal to the sum of the individual risks of the assets. This principle of diversification is crucial for effective risk management, as it ensures that combining different assets into a portfolio does not increase the overall risk disproportionately. Therefore, ES is often preferred over VaR for its ability to provide a more complete and reliable measure of risk, particularly in the context of extreme financial events.

Despite ES's superiority over VaR, it still has certain limitations in applicability to risk management, for instance, lack of robustness as discussed in \citet{cont2010robustness, kou2013external, he2022risk}. ES exemplifies a more general conflict between coherence and robustness. According to intuition, robust risk measures exhibit less sensitivity to outliers and mild misspecification of econometric models, which can lead to significant discrepancies in risk estimates. This lack of robustness in ES means that it can be disproportionately influenced by extreme observations or small errors in model specification, leading to unreliable risk assessments. \citet{cont2010robustness}   address this issue by proposing the RVaR, which aims to mitigate the sensitivity of risk measures to outliers and model misspecifications. The introduction of RVaR seeks to balance the coherence of a risk measure with the practical necessity of robustness, providing a more reliable tool for risk management in volatile financial environments.
\\
\textbf{3}. Given two significance levels, \(\alpha\) and \(\beta\), where \(0 < \alpha \leq \beta < 1\), the RVaR, denoted as \(\text{RVaR}^{\alpha,\beta}\), is defined as follows
\begin{equation}
\text{RVaR}^{\alpha,\beta}(X) = \begin{cases}
\frac{1}{\beta - \alpha} \int_{\alpha}^{\beta} \text{VaR}^u(X) \, du & \text{if } \alpha < \beta \\
-F_X^{-1}(\alpha) & \text{if } \alpha = \beta
\end{cases}
\end{equation}
RVaR bears a notable resemblance to ES, particularly as \(\alpha\) approaches zero. This similarity is accentuated by an alternative representation of RVaR \citep{wang2020risk}:
\begin{equation}\label{rvar}
\text{RVaR}^{\alpha,\beta}(X) = \frac{\beta \text{ES}^{\beta}(X) - \alpha \text{ES}^{\alpha}(X)}{\beta - \alpha}
\end{equation}
The validity of RVaR is further supported by the following inequality, which is derived from its properties:
\begin{equation}
\text{VaR}^{\alpha}(X) \geq \text{RVaR}^{\alpha,\beta}(X) \geq \text{VaR}^{\beta}(X), \quad \forall\; 0 < \alpha \leq \beta < 1
\end{equation}
This inequality indicates that RVaR is bounded above by \(\text{VaR}^{\alpha}(X)\) and below by \(\text{VaR}^{\beta}(X)\), ensuring its consistency and robustness as a risk measure.
\\
\subsection{\textnormal{\textit{Elicitability and scoring functions}}\label{3.4}}
Risk measures can be compared through elicitability property, facilitating the direct and theoretical comparison of alternative risk forecasting strategies \citep{gneiting2011making}. To illustrate elicitability, it is necessary to define scoring functions, which are maps \(S:\mathbb{R}^{2} \rightarrow (0,+\infty)\) that fulfill the following properties:
\begin{enumerate}
    \item \(S(x,y) = 0\) if and only if \(x = y\);
    \item \(y \mapsto S(x,y)\) is non-increasing for \(y < x\) and non-decreasing for \(y > x\), \(\forall x \in \mathbb{R}\);
    \item \(S(x,y)\) is continuous in \(y\), \(\forall x \in \mathbb{R}\).
\end{enumerate}
Scoring functions, also referred to as "loss functions", typically involve positive transformations of \(|x-y|\), \(x,y \in \mathbb{R}\). Zero-loss occurs only when there is no error, i.e., \(x = y\); otherwise, it increases with the distance between \(x\) and \(y\) and varies continuously. The composition of scoring functions is tailored to align with the distinct risk measures under assessment. Risk measures are considered elicitable if there exist some scoring functions \(S_{\rho} : \mathbb{R}^{2} \rightarrow \mathbb{R}_{+}\) such that:
\begin{equation}
\rho(X) = -\underset{y \in \mathbb{R}}{\arg\min} \mathbb{E}\left[S_{\rho}(X,y)\right], \quad \forall X \in \mathcal{G}
\end{equation}
In this scenario, the function \(S_{\rho}\) is considered consistent. Since risk is not directly observable, employing the actual risk as an argument for \(S_{\rho}(.,.)\) is not feasible. The function \(S_{\rho}\) assesses the error between the risk forecast and the return series, not using the true risk. According to \citet{fissler2021elicitability}, the VaR is elicitable under the following score function as
\begin{equation}
S_{\text{VaR}^\alpha}(x,y) = \alpha (x-y)^{+} + (1-\alpha) (x-y)^{-},
\end{equation}
While ES cannot be elicited individually, it is jointly elicitable with VaR, i.e., \(\rho(X):=\text{VaR}^{\alpha}, \text{ES}^{\alpha}\). According to \citet{gerlach2017semi}, the score function of ES under \(\mathbb{R}^{3}\), which is consistent, can be characterized as
\begin{equation}
S_{\text{ES}^{\alpha}}(x,y,z) = y(1_{x<y} - \alpha) - x1_{x<y}
 + e^z \left( z - y + \frac{1_{x<y}}{\alpha}(y-x) \right)
 - e^z + 1 - \log(1-\alpha)
\end{equation}
Similarly, RVaR cannot be elicited individually \citep{wang2020risk}, except through joint elicitability with the triplet \((\text{VaR}^{\alpha}, \text{VaR}^{\beta}, \text{RVaR}^{\alpha,\beta})\) \citep{fissler2021elicitability}. Hence, a strictly consistent scoring function with respect to \((\text{VaR}^{\alpha}, \text{VaR}^{\beta}, \text{RVaR}^{\alpha,\beta})\) can be defined as follows
\begin{equation}
\begin{split}
S_{\text{RVaR}^{\alpha,\beta}}(x,y,z,w) =  &y(1_{x<y} - \alpha) - x1_{x<y}+ z(1_{x<z} - \beta) - x1_{x<z} + (\beta-\alpha) \tanh((\beta-\alpha)w) \\&\left[ w + \frac{1}{\beta-\alpha} (S_{\text{VaR}^{\beta}}(x,z) - S_{\text{VaR}^{\alpha}}(x,y)) \right]
 - \log(\cosh((\alpha-\beta)w)) + 1 - \log(1-\alpha),
 \end{split}
\end{equation}
where \(S_{\text{VaR}^{(.)}}\) represents the scoring function of VaR defined in \eqref{var11}.
\\
\subsection{\textnormal{\textit{Model misspecification and legal robustness}}\label{3.5}} 
In the initial stages of the estimation process, it is essential to carefully choose a model. One frequent mistake is selecting a model that fails to capture the genuine risk dynamics in financial data. For instance, opting for a normal distribution with light tails despite observing heavy-tailed behavior in the data can result in substantial inaccuracies in risk. We believe that this source of model risk can be significantly mitigated by only considering models that are not rejected by backtesting. This viewpoint aligns with that of regulators, as regulatory authorities typically permit only those models for internal modeling that pass backtests. However, another issue persists: if regulatory authorities accept more than one model for internal modeling, it could create opportunities for regulatory arbitrage \citep{kellner2016quantifying}. In this context, we refer to the discussion in \citet{kou2013external}, which argues that two institutions with identical portfolios may use different internal models, both of which could gain regulatory approval. Nevertheless, these institutions should be mandated to hold equivalent or nearly equivalent amounts of regulatory capital because they manage the same portfolio. \citep{kellner2016quantifying} found that ES is more sensitive to regulatory arbitrage and parameter misspecification compared to VaR. The likelihood of regulatory arbitrage increases with the discrepancy in risk estimates among models. We applied this methodology to our portfolio data, which, to the best of our knowledge, has not been done before. We define the extent of regulatory arbitrage as the legal robustness, quantified by the mean absolute deviation among potential risk measure estimates relative to their average levels, as described below:
\begin{equation}
\begin{split}   &\text{LR}_{t+1}^{\text{VaR}_{\alpha}^{t+1}}=\frac{\frac{1}{n}\sum_{i=1}^{n}\left|\text{VaR}_{i,\alpha}^{t+1}-\overline{\text{VaR}}_{\alpha}^{t+1}\right|}{\overline{\text{VaR}}_{\alpha}^{t+1}},
\text{LR}_{t+1}^{\text{ES}_{\alpha}^{t+1}}=\frac{\frac{1}{n}\sum_{i=1}^{n}\left|\text{ES}_{i,\alpha}^{t+1}-\overline{\text{ES}}_{\alpha}^{t+1}\right|}{\overline{\text{ES}}_{\alpha}^{t+1}}, \text{and},\\&
\text{LR}_{t+1}^{\text{RVaR}_{\alpha,\beta}^{t+1}}=\frac{\frac{1}{n}\sum_{i=1}^{n}\left|\text{RVaR}_{i,\alpha,\beta}^{t+1}-\overline{\text{RVaR}}_{\alpha,\beta}^{t+1}\right|}{\overline{\text{RVaR}}_{\alpha,\beta}^{t+1}}
\end{split} 
\end{equation}
where $i=1,2,...,n,\overline{\text{VaR}}_{\alpha}^{t+1}=\frac{1}{n}\sum_{i=1}^{n}\text{VaR}_{i,\alpha}^{t+1},\overline{\text{ES}}_{\alpha}^{t+1}=\frac{1}{n}\sum_{i=1}^{n}\text{ES}_{i,\alpha}^{t+1}$, and $\overline{\text{RVaR}}_{\alpha,\beta}^{t+1}=\frac{1}{n}\sum_{i=1}^{n}\text{RVaR}_{i,\alpha,\beta}^{t+1}$ represent the average level of VaR, ES and RVaR estimates, respectively, across 
$n$ different models. The purpose of averaging these risk measures is to scale them uniformly, avoiding unfair comparisons between VaR, ES and RVaR. When this average level increases, it indicates more dispersed capital requirements among different banks or entities. This dispersion can potentially contribute to an unbalanced regulatory environment within the banking industry, where institutions facing similar risks may end up with varying capital requirements due to differences in their internal risk measurement models or methodologies.
\subsection{\textnormal{\textit{Methods for estimating risk measures}}\label{3.6}} 
Effective risk management is essential for guiding investment decisions and market strategies. This study explores the use of nonparametric, parametric, and Monte Carlo Simulation (MCS) methodologies to estimate the aforementioned risk measures. These approaches are chosen for their ability to provide robust and flexible estimation of risk metrics like VaR, ES, and RVaR, ensuring comprehensive risk assessment in decision-making processes.
\subsubsection{\textnormal{\textit{Non-parametric method}}}\label{3.6.1}
Non-parametric methodology for quantifying risk measures avoids strong assumptions about the distribution of returns. In this study, we focus on the Historical Simulation (HS) approach to compute VaR, ES, and RVaR. The HS method assumes that future returns will mirror historical patterns and that the probability distribution of returns remains unchanged. \citet{alexander2009market} argues that HS models simulate future projections based on past historical data, thereby generating a simulated distribution identical to the future returns distribution for a specified time horizon. To apply the HS approach, we begin by collecting return data for asset portfolios. Sorting and computing logarithmic accuracy establish the empirical performance distribution, from which VaR, ES, and RVaR can be extracted at desired confidence levels.
\\
For a portfolio with \( n \) assets, where \( r_{it} \) represents the return series for asset \( i \) at time \( t \), and \( w_{i} \) denotes the corresponding asset weights, the historical simulated returns for each time period \( t \) are calculated as
\[
r_{pt} = \sum_{i=1}^{n} w_{i} r_{it}.
\]
These simulated return series serve as the basis for estimating VaR, ES, and RVaR at specific confidence levels, ensuring a robust assessment of risk across the portfolio.
\subsubsection{\textnormal{\textit{Parametric method}}}\label{3.6.2}
In parametric risk measurement methods, risk is assessed by fitting probability distributions to the data and deriving estimates of VaR, ES, and RVaR from these fitted distributions. Parametric methods rely on specific assumptions about probability distributions, such as Normal, Student's $t$, Generalized Error Distribution, or other statistical distributions. In this study, all return series are modeled using the student's $t$ distribution, the curve of the return series is seen as leptokurtic justifying the adoption of the student's $t$ distribution. The formulas for calculating VaR based on the student's $t$ distribution are as follows
\begin{equation}\label{var3}
\text{VaR}_{t}= \mu_{t}+\sigma_{t}(T_{\nu}^{-1},P_{t-1})\sqrt{\left(\frac{\nu}{\nu-2}\right)},
\end{equation}
 where $\text{VaR}_{t}$ indicates the value at risk at time $t$, $P_{t-1}$ is the price of prior days of the returns, $\mu_{t}$ is the mean, $\sigma_{t}$ is the standard deviation of the returns at time $t$, and $T_{\nu}^{-1}$ is the quantile function of $t$ distribution with $\nu$ degree of freedom. To compute ES and RVaR, equations similar to those referenced as \eqref{es} and \eqref{rvar}, are utilized, following established mathematical frameworks in quantitative finance and risk management practices. These methodologies ensure rigorous assessment and management of risk based on the specified statistical assumptions.
\subsubsection{\textnormal{\textit{Monte Carlo Simulation (MCS) method}}}\label{3.6.3}
The MCS method is an empirical approach used to assess the prices or returns of financial assets, bypassing strict assumptions about their distribution. This method employs a non-linear return function to compute the risk of financial time series. MCS is based on generating random scenarios from possible states of the market, leveraging correlations and volatilities calculated by risk managers. Each simulation provides a feasible value for each asset in a portfolio at a specified future time horizon. The MCS approach can be implemented through the following steps. 
\begin{enumerate}
    \item \textit{Parameter Assessment:} Evaluate financial variables using appropriate methodologies.
    \item \textit{Simulation of distributions:} Generate simulated price scenarios for each asset based on assumed distributions and expected variations.
    \item \textit{Portfolio Valuation:} Calculate the value of the investment portfolio at time 
$t$ using simulated prices of financial assets.
    \item \textit{Building Portfolio Distribution:} Repeat steps (ii) and (iii) multiple times (e.g., 1000, 10,000, or 100,000 simulations) to build a distribution of portfolio values.
    \item \textit{Risk Measure Extraction:} Determine risk measures at a specified  level
$(1- \alpha)$ from the simulated distribution at time
$t$.
\end{enumerate}
Through these steps, MCS provides a robust framework for quantifying and managing financial risk by capturing the uncertainty and variability inherent in market conditions without relying on restrictive assumptions about asset returns.
\subsection{\textnormal{\textit{Copula modeling for dependence}}\label{3.7}}
In financial research, copulas are the mechanism that allows us to construct a multivariate distribution by separately defining the marginal distributions of financial assets and the copula functions. Through copula, it is possible to expose and understand the various fallacies associated with correlations among assets. Also, they are often used in a "black box" manner, and understanding the overall joint multivariate distribution can be challenging if marginals and copula are specified separately. Despite this, an understanding of copulas is imperative for risk management. As we shall use bivariate copula functions later, we begin by defining a copula on bivariate distributions.
\\
Let $X=(X_1, X_2)$ be a two-dimensional continuous random variable having CDF $F(x_1,x_2)$, their marginals CDF are defined as $G(x_1)$ and $H(x_1)$ ,respectively. The unique bivariate copula function can be defined as;
\begin{equation}\label{copula1}
        F(x_1,x_2)=C(G(x_1),H(x_2)).
\end{equation}
If $G(x_1)$ and $H(x_2)$ are continuous then $C(.)$ is uniquely identifiable, the $F(x_1,x_2)$ is the joint distribution function retaining $G(x_1)$ and $H(x_2)$ marginal. By following \citet{sklar1959fonctions}, the joint density function defined as
\begin{equation}\label{copula_pdf}
    {f(x,y)=c(G(x_1),H(x_2))g(x_1)h(x_2)}
\end{equation}
where $c$ is the copula density function with $g(x_1)$ and $h(x_1)$ marginal density functions, respectively. If the copula function is defined on continuous probability space, then the copula density function can be expressed as,
\begin{equation}
    c(u_1,u_2)=\frac{\partial^2C(u_1,u_2)}{\partial u_1 \partial u_2}
\end{equation}
 If the two-dimensional random variable $X=(X_1, X_2)$ with cumulative distribution function $ F(x_1,x_2)$ is on continuous space, their marginal distribution functions $G(x_1)$ and $H(x_1)$ are also definitive continuous and rigorously incremented, then the corresponding copula density function can be defined as;
\begin{equation}
    c(u_1,u_2)=\frac{f(G^{-1}(u_1, H^{-1}(u_2)))}{g(G^{-1}(u_1)h(H^{-1}(u_2))},
\end{equation}
where $f(x,y)$ is the joint density function of $F(x,y)$, $g(.)$ and $h(.)$ are marginal density functions. Our study examines the pairing relationship between cryptocurrencies and global financial indices. To enhance the efficiency of our modeling, we employ various copula families. The details of these copula families are given below subsequently. 
\subsubsection{\textnormal{\textit{Frank copula}}\label{3.7.3}}
The Frank copula function is defined as
\begin{equation}
    C(u_1,u_2)=\exp\left(-((-{ln} u_1)^\delta+(-{ln} u_2)^\delta)^{\frac{1}{\delta}}\right),
\end{equation}
where $\delta\geq1$.
\subsubsection{\textnormal{\textit{Gumbel copula}}\label{3.7.4}}
The Gumbel copula is expressed as
\begin{equation}
C(u_1,u_2)=-\frac{1}{\lambda}{ln}\left(1+\frac{(e^{-{\lambda}u{_1}_{-1}})(e^{-{\lambda u_2}}-1)}{(e^{-{\lambda}}-1)}\right),
\end{equation}
where $\lambda\neq 0$ and If $\lambda>0$ indicate the positive dependence, while $\lambda<0$ indicates negative dependence between financial variables.
\subsubsection{\textnormal{\textit{Joe copula}}\label{3.7.5}}
The Joe copula function is defined as
\begin{equation}
C(u_1,u_2)=1-\left[(1-u_1)^{\theta}+(1-u_2)^{\theta}-(1-u_1)^{\theta}(1-u_2)^{\theta}\right]^{\frac{1}{\theta}},
\end{equation}
where $1\leq\theta <\infty$ and Joe copula function performs like the Clayton copula and cannot acquire the negative dependence.
\subsubsection{\textnormal{\textit{Student's t copula}}\label{3.7.6}}
 The two-dimensional student's $\textit{t}$ copula function is defined as
 \begin{equation}
   C(u_1,u_2)=int_{-\infty}^{t_\nu ^{-1}(u_1)}\int_{-\infty}^{t_\nu ^{-1}(u_2)}\frac{1}{2\pi\sqrt{1-\theta^{2}}}
   \left(1+\frac{s^{2}-2\theta st+t^{2}}{\nu(1-\theta^{2})}\right)^{\frac{\nu+2}{\nu}} ds dt,
 \end{equation}
 where $t_\nu ^{-1}(.)$ indicate the inverse function of the univariate student's $\textit{t}$ distribution function $t_\nu(.)$ with $\nu$ degree of freedom. The function is symmetrical like a normal distribution and estimates the distribution lower and upper tail dependence. In addition to providing benchmarks for ordering dependence, it also significantly affects the increase or decrease of one variable about another. It is easy to find further details about these copula families in literature, therefore they are not discussed here in detail. In addition, we will estimate the parameters of the above copula function through MLE.
 \subsubsection{\textnormal{\textit{Simulation through copulas}}\label{3.7.7}}

In this section, we aim to explain about simulation of financial variables using copula functions within the MCS framework. This involves generating pairs of observations 
$(u_1, u_2)$
 from uniformly distributed random variables 
$(U_1, U_2)$ on the interval 
$(0, 1)$ using conditional sampling along with given specific copula parameters. The overall conditional distribution is defined as
\[
c_{u_1}(u_2) = \Pr(U_2 \leq u_2 \,|\, U_1 = u_1),
\]
and the conditional distribution of 
$U_2$ given $U_1$ can be derived as 

\[
c_{u_1}(u_2) = \Pr(U_2 \leq u_2 \,|\, U_1 = u_1)
=\lim_{\Delta u_1\to 0}\frac{c(u_1+\Delta u_1, u_2)-c(u_1,u_2)}{\Delta u_1}=\frac{\partial C}{\partial u_1}=c_{u_1}(u_2)
\] 

Our copula-based simulation for this study follows the following steps.
\begin{enumerate}
\item Generate two independent random variables \(U_1\) and \(U_2\) from the uniform distribution on \([0,1]\):
        \[
        U_1 \sim \mathcal{U}(0,1), \quad U_2 \sim \mathcal{U}(0,1).
        \]
\item Calculate the conditional distribution function of the copula. Let \(C(u_1, u_2; \delta)\) be a copula function parameterized by \(\delta\). The conditional distribution function is defined as:
        \[
        C(u_2|u_1) = \frac{\partial C(u_1, u_2; \delta)}{\partial u_1}.
        \]
 Compute the inverse of the conditional distribution function at \(U_2\):
        \[
        U_2 = C^{-1}(u_2 | U_1),
        \]
        where \(C^{-1}(\cdot | u_1)\) is the inverse of the conditional copula function with respect to \(u_2\).
 \item The pairs \((U_1, U_2)\) are now dependent and uniformly distributed on \([0,1]^2\) according to the copula function \(C\). Transform these uniformly distributed variables into the desired marginals. Let \(G(x_1)\) and \(H(x_2)\) be the marginal distribution functions of the random variables \(X_1\) and \(X_2\), respectively. Simulated observations from the joint distribution function \(F(x_1, x_2) = C(G(x_1), H(x_2))\) can be obtained by:
        \[
        (X_1, X_2) = \left(G^{-1}(U_1), H^{-1}(U_2)\right),
        \]
        where \(G^{-1}\) and \(H^{-1}\) are the inverse functions of the marginal distribution functions \(G\) and \(H\), respectively.
\end{enumerate}

\subsection{\textnormal{\textit{Spillover analysis paradigm}}\label{3.8}}

The volatility of the financial markets generally increases sharply in crises and spills over to other markets as well.  In order to provide early warning systems for emerging crises and to monitor the progress of existing crises, it would naturally be desirable to be able to measure and monitor such spillovers. Based on this ground, \citet{diebold2012better} proposed the spillover index by exploiting forecast error variance analysis (FEVDs) of the generalized auto-regressive (GVAR). Given
$n$ process variables, $x_t=(x_{t1},x_{t2},.....x_{tn})$ then Structural Vector Auto-regression Model (SVAR) having order of $(p)$ can be expressed as follows
\[
\Phi(L)x_t = \mu_t,
\]
where, \( t = 1, 2, \ldots, T \). In the above expression, \( \Phi(L) = \sum_{h=0}^{p} \Phi_h L^h \) is an \( n \times n \) lag polynomial matrix of order \( p \), and \( \mu_t \) represents the error term (white noise) with a covariance matrix \( \Sigma \). Provided that the roots of the determinant \( |\Phi(z)| \) lie outside the unit circle, the Vector Autoregressive model can be rewritten as:
\[
x_t = \varphi(L)\mu_t
\]
According to \citet{diebold2012better}, the coefficients \(\varphi(L)\) correspond to the covariance matrix coefficients. The H-step generalized forecast error variance decomposition (FEVD) is given by:
\begin{equation*}
(\gamma_{H})_{j,k} = \frac{\sigma_{kk}^{-1} \sum_{h=0}^{H} \left( (\Phi_h \Sigma)_{j,k} \right)^2}{\sum_{h=0}^{H} (\Phi_h \Sigma \Phi'_h)_{j,j}},
\end{equation*}
where \( \Phi_h \) represents the \( n \times n \) coefficient matrix at lag \( h \) and \( \sigma_{kk} = (\Sigma)_{kk} \). The term \( (\gamma_{H})_{j,k} \) indicates the influence of the \( k \)-th variable on the forecast error variance of the \( j \)-th variable. The normalization of the decomposition matrix elements is achieved through:
\[
(\Tilde{\gamma}_{H})_{j,k} = \frac{(\gamma_{H})_{j,k}}{\sum_{k=0}^{n} (\Tilde{\gamma}_{H})_{j,k}}, \quad \text{where} \quad \sum_{k=0}^{n} (\Tilde{\gamma}_{H})_{j,k} = 1 \quad
\text{and} \quad \sum_{j,k=1}^{n} (\Tilde{\gamma}_{H})_{j,k} = N
\]
The connectedness measure is then derived as:
\begin{equation*}
T_H = \frac{\sum_{j \neq k} (\Tilde{\gamma}_{H})_{j,k}}{\sum (\Tilde{\gamma}_{H})_{j,k}} \times 100 = 100 \left( 1 - \frac{B_r\{\Tilde{\gamma}_H\}}{\sum (\Tilde{\gamma}_{H})_{j,k}} \right)
\end{equation*}
where \( B_r\{.\} \) is the trace operator.

\section{Empirical investigations}\label{applications}
\subsection{\textnormal{\textit{Data}}\label{4.1}}
\begin{figure*}
  \centering
  \includegraphics[width=162mm, height=105mm]{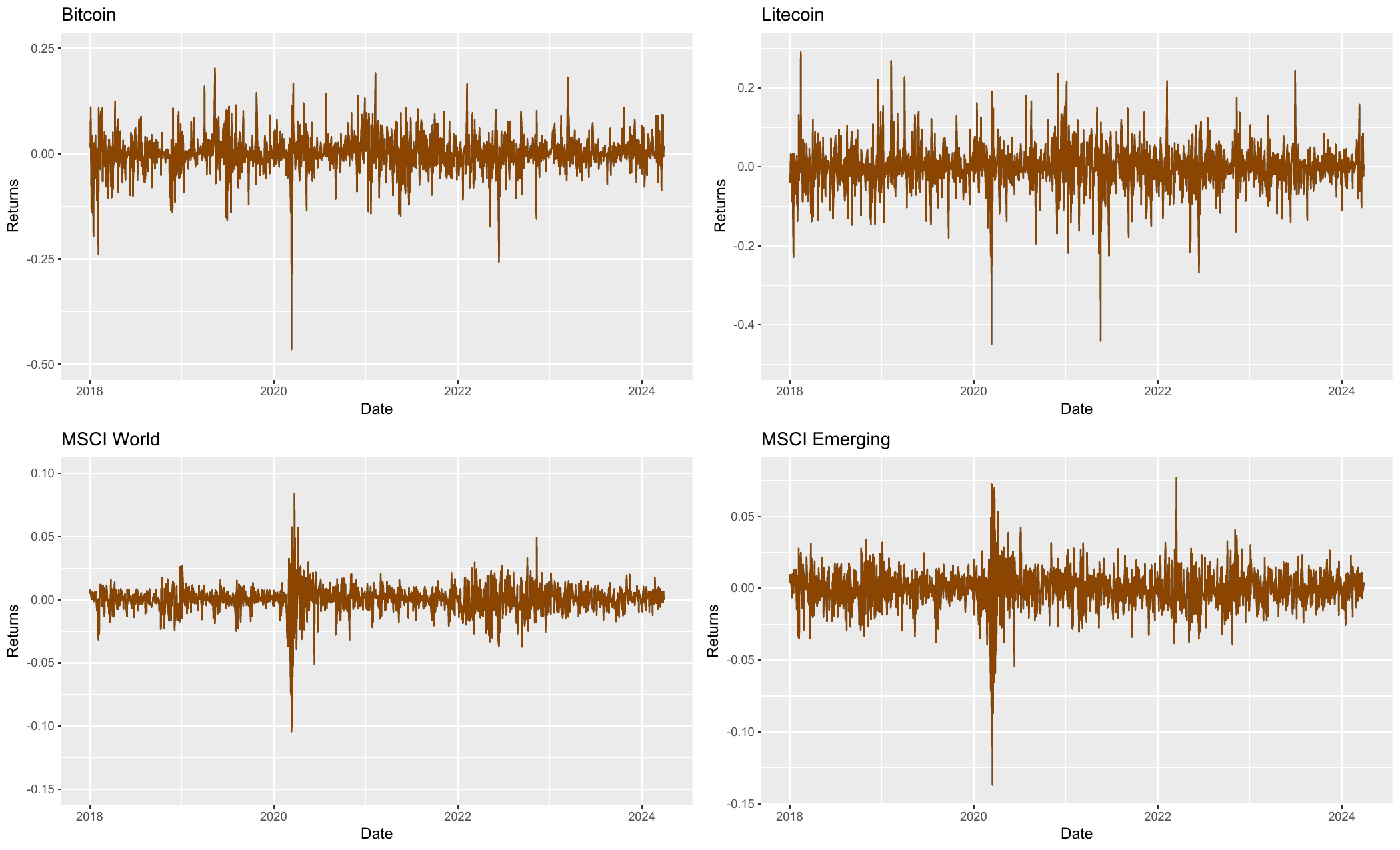}  
\caption{ Bitcoin, Litecoin, MSCI World, and MSCI Emerging returns.} 
\label{fig:1}
\end{figure*}

In empirical applications, we collected the daily closing prices data of Bitcoin, Litecoin, MSCI World Index, and MSCI Emerging Markets Index, spanning from January 1, 2018, to March 31, 2024. The data was obtained from \url{http://finance.yahoo.com/}. The lengths of the datasets are not identical due to the different operational days of these markets. Before modeling, we adjusted the time intervals accordingly. The observed series were found to be non-stationary. To transform them into stationary series, we computed the logarithmic returns of the non-stationary observed series as follows:
\begin{equation}            
    R_t=\text{ln}\left(\frac{P_{t}}{P_{t-1}}\right)
\end{equation}
Here, $R_t$ represents the rate of returns sequence, and $P_t$ and $P_{t-1}$ are the closing prices of the markets at time $t$ and $t-1$, respectively. Figure \ref{fig:1} illustrates the return series of the four markets: Bitcoin, Litecoin, MSCI World, and MSCI Emerging Markets.
We further assessed the stationarity of each series using the Augmented Dickey-Fuller (ADF) and Phillips-Perron (PP) tests. The results, detailed in Table~\ref{tab:stationarity} confirm that all series are stationary. 
\begin{table*}[htbp]
  \centering
  \caption{ADF and PP test outcomes and $p$-values of each return series}
  \begin{tabular}{lcccccc}
    \toprule
     & \multicolumn{2}{c}{ADF Test} & \multicolumn{2}{c}{PP Test}& \multicolumn{2}{c}{Decision} \\
    \cmidrule(lr){2-3} \cmidrule(lr){4-5}
             & Statistics & $p$-Value & Statistics &$p$-Value &\\
    \midrule
    Bitcoin         &-10.631   &0.0000 &-1744.40 & 0.0000& Stable\\
    Litecoin      &-10.911   &0.0000 &-1663.80  & 0.0000& Stable\\
    MSCI\_W      &-11.070    &0.0000 &-1733.00  & 0.0000& Stable\\
    MSCI\_EM     &-11.609   &0.0000 &-1930.10  & 0.0000& Stable\\
    \bottomrule
  \end{tabular}
    \label{tab:stationarity}
\end{table*}
\\
Additionally, Table~\ref{tab:summary} presents the summary statistics of the transformed series along with the results of the normality test Jarque-Bera. The negative skewness values and kurtosis values greater than 3 indicate deviations from normality and the presence of high peaks. Furthermore, the Jarque-Bera test results confirm that the data do not follow a normal distribution.

\begin{table*}[H]
    \centering
    \caption{Descriptive statistics on return series}
    \begin{tabular}{p{2cm} p{1.5cm} p{1.5cm} p{1.4cm} p{1.4cm} p{1.4cm} p{1.5cm}p{2cm}}
    \toprule
    & Minimum & Maximum & Mean &Std & Skewness & Kurtosis& Jarque-Bera \\
    \cmidrule{1-8}
    Bitcoin &-0.46473 &0.20305&0.00099&0.04414 &-0.97267  &10.94992  &8075.610***\\
    Litecoin &-0.44906 &0.29059 &-0.00063 &0.05810 &-0.56148 & 6.75910 &3065.215***\\
    MSCI\_W &-0.10440  &0.08406 &0.00030 &0.01089 &-1.04150 & 15.42483 &15817.840*** \\
    MSCI\_EM &-0.13691& 0.07693 &-0.00008 &0.01410 &-0.93871 &11.01853 &8157.062*** \\
    \bottomrule
    \end{tabular}
    \begin{tablenotes}
    \item Note: *** represent 1\% significance level.  
    \end{tablenotes}
    \label{tab:summary}
\end{table*}
The analyses conducted indicated the presence of volatility clustering effects in all return series. 
Consequently, our study aims to utilize various GARCH models to model the volatility of each series. Before applying the models, we used the Auto-Regressive Conditional Heteroscedasticity Lagrange Multiplier (ARCH-LM) test to check for the presence of ARCH effects in the series. Additionally, we employed the Ljung-Box test to determine whether autoregressive terms should be included in the conditional mean equations. The results are summarized in Table \ref{tab:ARCH}. The p-values associated with the ARCH-LM test indicate that all the series exhibit ARCH effects, providing a basis for applying GARCH models. In addition, the Ljung-Box test shows that all the series have no auto-correlation except the MSCI\_EM series. However, emerging markets are less efficient and more liable to political and economic instability than developed markets. These test results indicate the need to apply the GARCH model to capture the volatility dynamics of the return series.

\begin{table}[htbpp]
  \centering
  \caption{ARCH-LM and Ljung-Box test results and $p$-values of each return series}
  \begin{tabular}{lccccc}
    \toprule
     & \multicolumn{2}{c}{ARCH-LM Test} & \multicolumn{2}{c}{Ljung-Box Test} \\
    \cmidrule(lr){2-3} \cmidrule(lr){4-5}
            & Statistics & $p$-Value & Statistics &$p$-Value \\
    \midrule
    Bitcoin         &275.9957  &0.0000  &2.6555   &0.1032\\
    Litecoin        &112.7031  &0.0000  &2.9421   &0.0863\\
    MSCI\_W      &490.2466  & 0.0000  &1.2492  &0.2637 \\
    MSCI\_EM     & 126.9376 & 0.0000  &14.8293  &0.0000\\
    \bottomrule
  \end{tabular}
    \label{tab:ARCH}
\end{table}

\subsection{\textnormal{\textit{GARCH models fitting}}\label{4.4}}
In order to capture the volatility in the return series, the family of the GARCH models introduced in subsection \ref{Volatility_model} is fitted and assess their optimal fitting for our return series through different assessment (Akaike's Information Criterion (AIC), Bayesian information criterion (BIC), mean absolute error (MAE), and root mean square error (RMSE)) criterion. The results of assessment measures are given in Table~\ref{tab:GARCH_asses}.
\begin{table*}[H]
    \centering
    \caption{AIC, BIC, MAE, and RMSE values of each GARCH model. The values in bold represent the optimal fit.  }
    \begin{tabular}{p{2.8cm}p{2.5cm} p{2.3cm} p{2.3cm} p{2.3cm} p{1.2cm}}
        \toprule
        Variables& Models& AIC& BIC & MAE &RMSE  \\
        \cmidrule{1-6}
             &\textit{s}GARCH&-3.7459 &-3.7185 &0.0292 &0.0441\\
        Bitcoin&\textit{e}GARCH& $\bold{-3.7570}$ & $\bold{-3.7262}$&$\bold{0.0291}$ &$\bold{ 0.0440}$\\
            &\textit{gjr}GARCH&-3.7450 &-3.7142 & 0.0293 &0.0443 \\
            \cmidrule{2-6}
                         &\textit{s}GARCH&-3.1030  &-3.0586 & 0.0396 &0.0581 \\
        Litecoin&\textit{e}GARCH&$\bold{-3.1105}$ & $\bold{-3.0627}$& $\bold{ 0.0394}$& $\bold{0.0580}$\\
            &\textit{gjr}GARCH&-3.1021 &-3.0542 &0.0395 &  0.0581\\
            \cmidrule{2-6}
                        &\textit{s}GARCH& -6.7744 & -6.7129 & 0.0071 & 0.0109\\
        MSCI\_W&\textit{e}GARCH& $\bold{-6.7811}$&$\bold{-6.7196}$& $\bold{0.0070}$ & $\bold{ 0.0106}$\\
            &\textit{gjr}GARCH&-6.7744 & -6.7129 &0.0071 & 0.0109\\
              \cmidrule{2-6}
            &\textit{s}GARCH&-5.9954 &-5.9441& 0.0097 & 0.0138\\
        MSCI\_EM&\textit{e}GARCH&$\bold{-6.0007}$&$\bold{-5.9460}$& $\bold{0.0096}$&$\bold{0.0136}$\\
            &\textit{gjr}GARCH&-5.9997 &-5.9450&0.0098 & 0.0139\\
        \bottomrule
    \end{tabular}
    \label{tab:GARCH_asses}
\end{table*}

In all return series, the \texttt{eGARCH} model show the best fitting with the lowest values of assessment criteria. As we have seen in Table \ref{tab:summary}, each return series often exhibits a leptokurtic distribution with thicker tail. Given the evidence from characteristics of the return series, it is indeed difficult to assume that the distribution of residuals approaches a Gaussian distribution. This observation necessitates the use of more sophisticated distribution that can more accurately capture the inherent characteristics of the return series. In this paper, we assume that the residuals follow a $t$ distribution, where the shape parameter represents the degrees of freedom of the $t$ distribution. The $t$ distribution can better capture the sharp peaks and thick tails often observed in the return series, compared to the normal distribution. The results of the fitted \texttt{eGARCH} model are summarized in Table~\ref{tab:5} 
\begin{table*}[H]
    \centering
    \caption{Parameters of optimal GARCH model for each return sequence}
    
    \begin{tabular}{p{2cm}p{2.3cm}p{2.3cm} p{2.3cm} p{2.3cm} p{2.3cm}}
        \toprule
       Bitcoin &mu   &ma1 &  ma2 & ma3 & omega  \\
               &0.000714 &-0.038610* & 0.033771  &-0.009941 &-0.113776*** \\
        &  alpha1& beta1 &gamma1& shape &\\
       &0.016817 &0.981025***&0.239822***&2.686661***& \\
\cmidrule{2-6}
      Litecoin &mu   &ar1 &  ar2 & ar3 & ar4 \\
               & 0.000335 & -0.968232*** &0.345071*** & 0.488784***& -0.010119   \\
        &  ar5 & ma1& ma2& ma3  &omega\\
        &  0.002078 & 0.926076*** & -0.383874*** &-0.491745*** &-0.287009*** \\
         &alpha1 &beta1& gamma1   & shape \\
         & 0.006581  & 0.948171***&0.238400***  &2.997667***\\
\cmidrule{2-6}
       MSCI\_W &mu   &ar1 &  ar2 & ar3 & ar4 \\
               & 0.000893*** & -0.685346*** & 0.738178*** &-0.111056***&-1.132095***\\
        &  ar5 & ar6& ar7 & ma1 &ma2 \\
        & -0.287999*** &0.015104***  & -0.028022*** & 0.794591*** & -0.665112***  \\
         &ma3 &ma4 &ma5 &omega & alpha1\\
         &0.000974 &1.127751*** & 0.425648*** & 0.000002* &0.164138***\\
         &beta1& shape & &\\
         &0.821421*** &6.753132*** & & \\
\cmidrule{2-6}
       MSCI\_EM &mu   &ar1 &  ar2 & ar3 & ar4 \\
               &0.000074 &-0.660752***  &-0.054781*   & -0.509175*** & 0.131320***  \\
        &  ar5 & ma1& ma2 & ma3 &ma4 \\
        &0.767937*** & 0.623901*** &0.020878  & 0.492900***  &-0.164865***\\
         &ma5 &omega &alpha1 &beta1 & gamma1\\
         &-0.785671***  &   -0.339879*** &-0.088470***  &   0.961804*** &0.155869*** \\
         &shape &&&&\\
         & 9.933018***&&&&\\
        \bottomrule
        \end{tabular}
    \begin{tablenotes}
      \item Note: ***, **, and * symbolizes significance levels at 1\%, 5\%, and 10\%, respectively.
    \end{tablenotes}
    \label{tab:5}
\end{table*}
The findings of Table~\ref{tab:5} reveals that in the \texttt{eGARCH} model for each return series, both the 
alpha1 and beta1 coefficients are highly significant. Importantly, these coefficients satisfy the stability condition where alpha1 +beta1 $< $1 as similar to \citet{zhao2023extreme}. This indicates that the GARCH model is robust and does not exhibit explosive behavior in volatility dynamics. Furthermore, with alpha1 +beta1 $< $1 or being close to 1, it suggests that changes in volatility tend to persist over time. This persistence is a notable characteristic in financial returns, illustrating how past shocks and volatility levels influence current and future volatility levels. The results of the ARCH-LM test for both the residuals and their squares indicate the absence of a significant ARCH effect. This suggests that the \texttt{eGARCH} model employed effectively mitigates any remaining autocorrelation in the original sequence. Overall, the model demonstrates a good fit to the data.

\subsection{\textnormal{\textit{Fitting the EVT models}}\label{4.5}}
\begin{figure*}
  \centering
  \includegraphics[width=162mm, height=140mm]{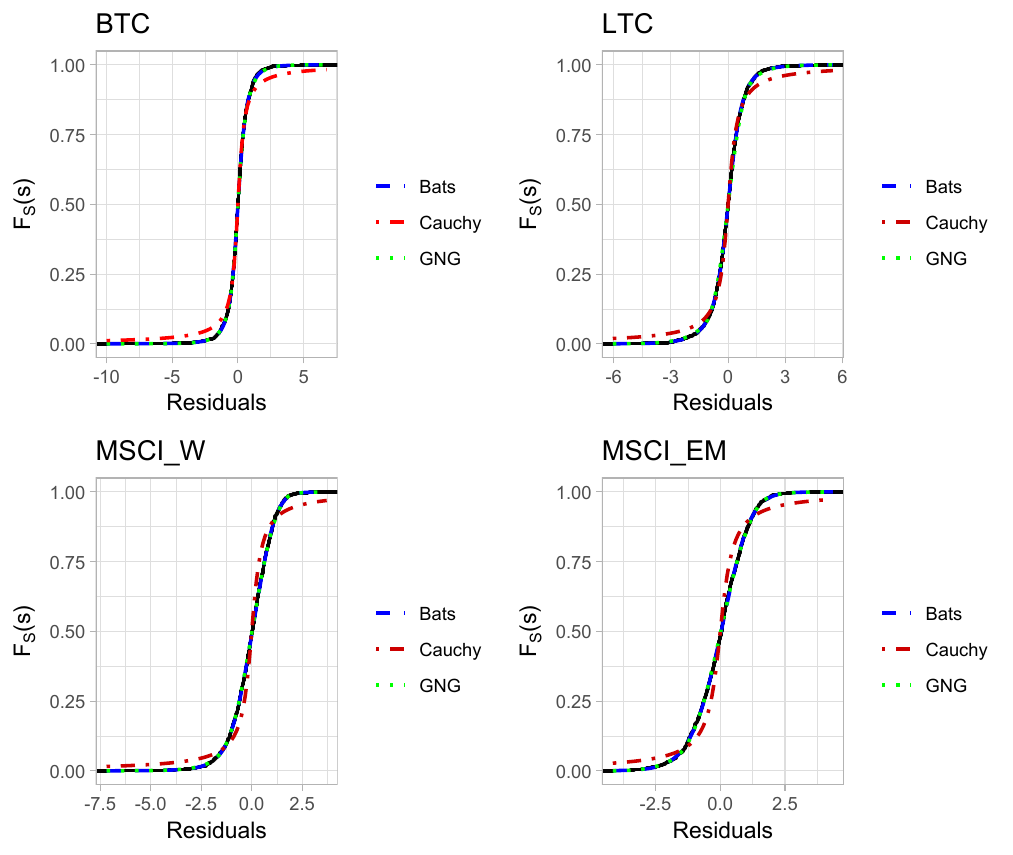}  
\caption{CDF charts for Bitcoin, Litecoin, MSCI World, and MSCI Emerging Markets} 
\label{fig:2}
\end{figure*}

\begin{figure*}
     \centering
     \begin{subfigure}[b]{0.49\textwidth}
         \centering
         \includegraphics[width=75mm, height=71.5mm]{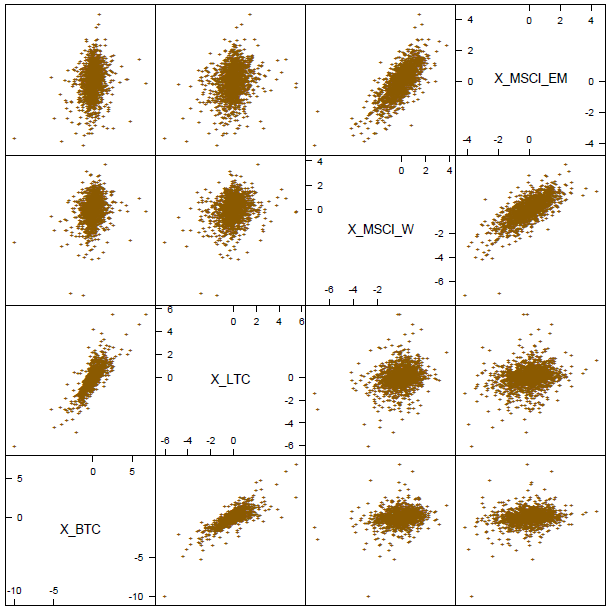}
     \end{subfigure}
     \hfill
     \begin{subfigure}[b]{0.49\textwidth}
         \centering
         \includegraphics[width=75mm, height=71.5mm]{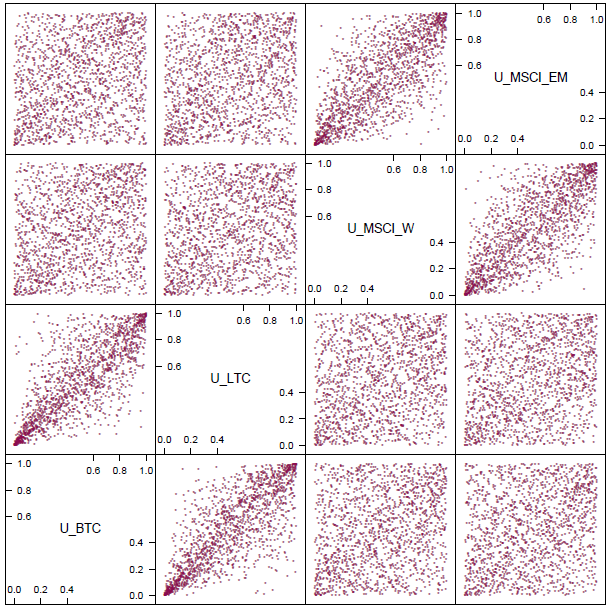}
     \end{subfigure}
     \hfill
        \caption{Standardized residual data series (left); Uniformly transformed data series (right)}
        \label{fig:3}
\end{figure*}

We fitted the models introduced in Section~\ref{Riskmodel} to the marginal return series using Maximum Likelihood Estimation (MLE) and assessed their optimal fitting through various statistical criteria, namely the Bayesian Information Criterion (BIC), Cramer-von Mises (CVM), and Anderson-Darling (AD) tests. The estimated parameters and goodness-of-fit results are presented in Table~\ref{tab:6}. The lowest values, highlighted in bold, indicate that the BATs model is the most appropriate for the marginal return series. Additionally, Figure~\ref{fig:2} illustrates the graphical behavior of the estimated cumulative distribution functions (CDFs) of all considered models. The BATs model provides the best fit for the marginal series, with the GNG model closely following, whereas the Cauchy model consistently performs the worst.
It is noteworthy that the application of the BATs model in financial research appears to be unprecedented, as this study is the first to do so.
\\
Given these findings, we adopt the BATs CDF in subsequent as the marginal distribution to observe the dependence among all four return series using copula models. Figure~\ref{fig:3} displays the standardized residuals and their uniform data.
\begin{table*}[hh]
    \centering
    \caption{Estimated parameters of both tail flexible models and their goodness-of-fit checks.  }
    \begin{tabular}{p{2cm} p{1.6cm} p{4.8cm} p{1.8cm} p{1.8cm} p{1.38cm}}
        \toprule
        Variables & Models & Parameters & BIC & CVM-test& AD-test\\
        \cmidrule{1-6} 
        & GNG &$\lambda_1= 0.2012$,$\delta_1=0.4644$ & 3692.907&0.0191&0.1990\\
        && $\xi_1=0.2001$,$\mu=1.3873$,&&[0.9978]&[0.9907] \\
        && $\sigma=24.0591$,$\lambda_2=0.1328$,&&&\\
        &&$\delta_2=0.5436$, $\xi_2=0.0838$&&&\\
        Bitcoin&BATs&$\kappa_{0}=0.2331$,$\tau_{0}=0.4153$,&$\bold{3676.064}$&$\bold{0.0121}$&$\bold{0.1203}$\\
        & &$\varphi_{0}=-0.2042$,$\kappa_{1}=0.1423$,&&$\bold{[0.9999]}$&$\bold{[0.9998]}$\\
        & &$\tau_{1}=0.4655$,$\varphi_{1}= 0.2503$,&&&\\
        &&$\nu=0.9514$&&&\\
        &Cauchy &$\lambda=0.0024$, $\delta=0.3615$ &3795.203  & 0.5272&7.8975\\
        &&&&[0.0339]&[0.0001]\\
    
    \cmidrule{2-6}
        & GNG &$\lambda_1= 1.3286$,$\delta_1=1.5704$, & 3969.804&0.2956&1.5421\\
        &&$\xi_1=-0.1262$,$\mu=0.0145$,&&[0.1390]&[ 0.1667]\\
        &&$\sigma=0.6446$,$\lambda_2= 1.3300$,&&&\\
        &&$\delta_2=0.7259$, $\xi_2=0.0688$&&&\\
        
Litecoin&BATs&$\kappa_{0}=-0.0793$,$\tau_{0}=2.2081$,&$\bold{3933.706}$&$\bold{ 0.0120}$&$\bold{0.0970}$\\
        & &$\varphi_{0}=1.0235$,$\kappa_{1}=0.1661$,&&$\bold{[1.0000]}$&$\bold{[1.0000]}$\\
        & &$\tau_{1}=1.4454$,$\varphi_{1}=-0.9430$,&&&\\
        &&$\nu=2.1415$&&&\\
        &Cauchy &$\lambda=0.0096$, $\delta= 0.4068 $ & 4092.529  & 0.6882&9.5997\\
        &&&&[0.0030]&[0.0000]\\
        \cmidrule{2-6}
& GNG &$\lambda_1= 1.6472$,$\delta_1=0.7167$, & 4444.903 & 0.0319&0.1975\\
        &&$\xi_1=0.1168$,$\mu=0.0623$,&&[0.9697]&[0.9911] \\
        &&$\sigma=0.9161$,$\lambda_2=1.6027$,&&&\\
        &&$\delta_2=0.2978$, $\xi_2=0.2210$&&&\\
        
        MSCI\_W&BATs&$\kappa_{0}= 1.3528$,$\tau_{0}=0.5599$,&$\bold{4430.27}$&$\bold{0.0169}$&$\bold{ 0.1142}$\\
        & &$\varphi_{0}=-0.2895$,$\kappa_{1}=0.6363$,&&$\bold{[0.9991]}$&$\bold{[0.9999]}$\\
        & &$\tau_{1}=0.6549$,$\varphi_{1}=0.5134$,&&&\\
        &&$\nu=5.1534$&&&\\
        &Cauchy &$\lambda=0.0494$, $\delta=0.5620$ & 4831.426 &1.7610 &19.9467 \\
        &&&&[0.0000]&[0.0000]\\
        \cmidrule{2-6}
& GNG &$\lambda_1=1.8821$,$\delta_1=0.7579$, &4507.274 &0.0655 &0.3207\\
        &&$\xi_1=-0.2169$,$\mu=0.0617$,&&[ 0.7794]&[0.9218] \\
        &&$\sigma= 0.9436$,$\lambda_2= 1.4516$,&&&\\
        &&$\delta_2=0.3806$, $\xi_2=0.1342$&&&\\
        
MSCI\_EM &BATs&$\kappa_{0}=0.7143$,$\tau_{0}=1.0574$,&$\bold{4496.994}$&$\bold{ 0.0457}$&$\bold{0.2695}$\\
        & &$\varphi_{0}=-0.1306$,$\kappa_{1}=0.6431$,&&$\bold{[0.9013]}$&$\bold{[0.9591]}$\\
        & &$\tau_{1}=0.8756$,$\varphi_{1}=0.1972$,&&&\\
        &&$\nu=5.9760$&&&\\
        &Cauchy &$\lambda=0.0512$, $\delta=0.5687$ &4892.514 &1.6112&19.0199\\
        &&&&[0.0000]&[0.0000]\\
        \bottomrule
    \end{tabular}
     \begin{tablenotes}
      \item Note: The values in brackets indicate the
$p$-values associated with the goodness of fit tests.
    \end{tablenotes}
    \label{tab:6}
\end{table*}

\subsection{\textnormal{\textit{Correlation vs dependence through copula }}\label{4.6}}
\begin{figure*}
  \centering
  \includegraphics[width=160mm, height=76mm]{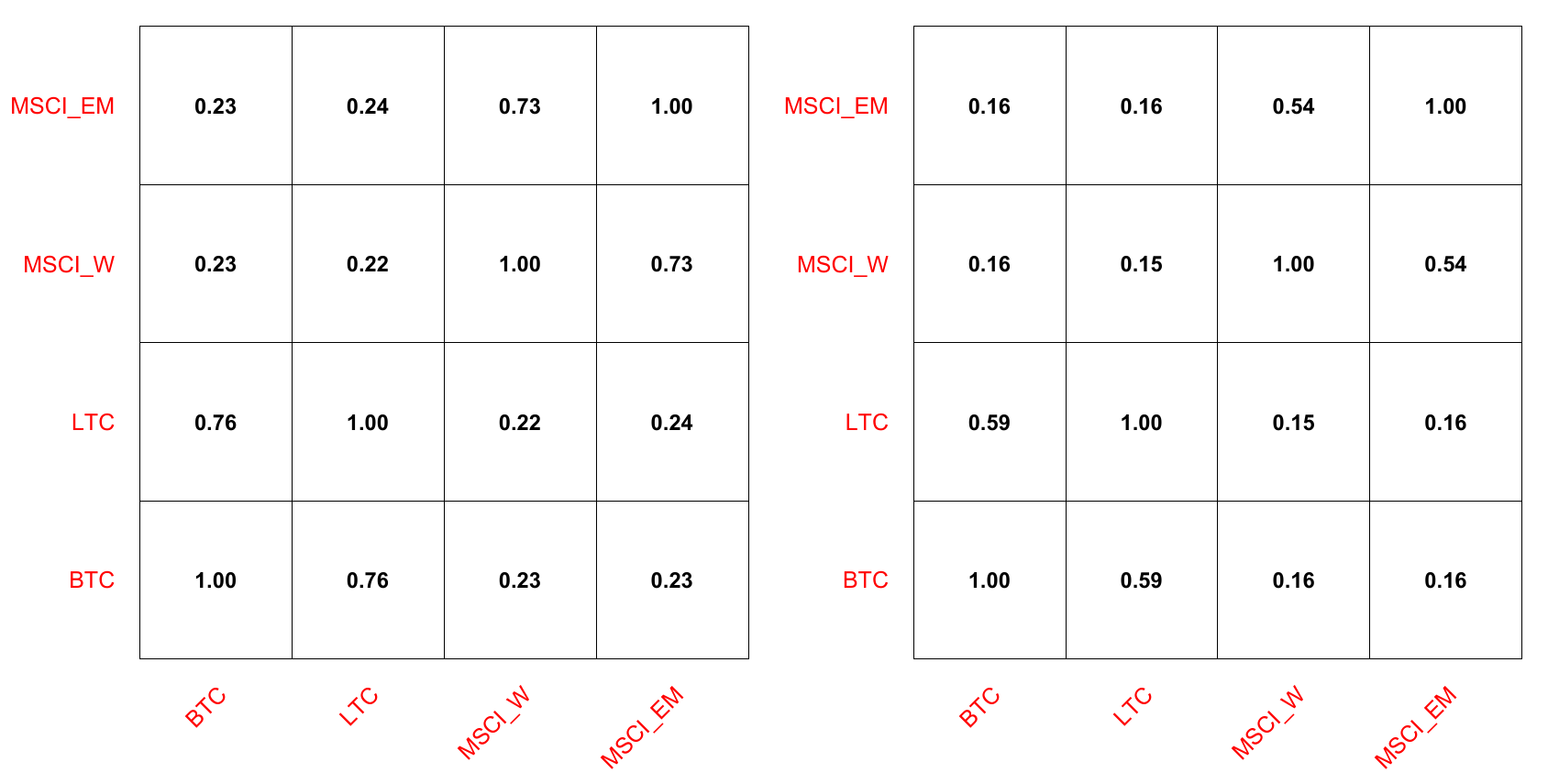}  
\caption{ The results of Spearman's (left) and Kendall's (right) tests.} 
\label{fig:4}
\end{figure*}

We estimated the correlation between the return series using Spearman’s rank correlation and Kendall’s tau coefficient. The dependence values from both tests are presented in Table~\ref{tab:7} and depicted in Figure~\ref{fig:4}. In Table~\ref{tab:7}, Spearman’s test shows a significant correlation of 0.7647 between Bitcoin and Litecoin, indicating substantial dependence or a strong relationship between these two cryptocurrencies. The results also demonstrate a low association between the cryptocurrency series and the MSCI World and MSCI Emerging Markets asset series. Similar to cryptocurrencies, when looking at MSCI World and MSCI Emerging Markets, we found a strong relationship among them. Kendall's test also shows a moderate correlation between the two cryptocurrencies and a weak relationship between cryptocurrencies and the MSCI World and MSCI Emerging Markets. 
\begin{table*}[htbpp]
  \centering
  \caption{Outcomes of Spearman's and Kendall's tau correlation coefficients tests}
\begin{tabular}{p{1.8cm}p{1.1cm}p{1.1cm} p{1.1cm} p{1.8cm} p{1.1cm}p{1.1cm}p{1.1cm}p{1.4cm}}
    \toprule
     & \multicolumn{4}{c}{Spearman's Test} & \multicolumn{4}{c}{Kendall's Test} \\
    \cmidrule(lr){2-5} \cmidrule(lr){6-9}
            & Bitcoin  & Litecoin & MSCI\_W &MSCI\_EM & Bitcoin  & Litecoin & MSCI\_W &MSCI\_EM\\
    \midrule
    Bitcoin   &1   &0.7647 & 0.2309 &0.2286   &1 & 0.5857&0.1572 &0.1558 \\
    Litecoin  &0.7647 &1    & 0.2192 &0.2402   &0.5857&1 &0.1471 &0.1619\\
    MSCI\_W  &0.2309 & 0.2192 &1   & 0.7278   & 0.1572& 0.1471& 1&0.5392 \\
    MSCI\_EM  &0.2286 &  0.2402  & 0.7278 &1  &0.1558 &0.1619 &0.5392 &1\\
    \bottomrule
  \end{tabular}
    \label{tab:7}
\end{table*}

On the other hand, we used different copula families to observe joint dependence among cryptocurrencies and global financial indices. This is crucial for calculating the copula-based VaR, ES, and RVaR. In the copula setting, the BATs are used as the marginal distribution of each return series, and bivariate copula models are fitted through MLE. The results are presented in Table~\ref{tab:8}. The findings in Table~\ref{tab:8} indicate a significant dependence between the pairs of assets. The coefficients for the Gumbel and Joe copulas reflect heightened turbulence during recessionary periods. In contrast, the coefficients for the Student's $t$ copula suggest that dependence is primarily concentrated in the tails of the distribution rather than its central points. We observed that the joint dependence estimates obtained through Student's $t$ copula appear very near to the original dependence structure of the return series. 
\begin{table*}
  \centering
  \caption{Results of the copula parameters for each paired portfolio}
  \begin{tabular}{p{4.9cm}p{2.3cm}p{2.3cm} p{2.3cm} p{2cm}}
    \toprule
       Portfolios &Frank & Gumbel & Joe & Student's t \\
        \cmidrule{1-5}
    $\bold{Bitcoin-MSCI\_W}$ & $\theta=1.459$   & $\theta=1.144$  & $\theta=1.145$   &$\theta= 0.237$,\\ 
    &[0.009] &[0.000] & [0.009]&$\nu=11.396$\\
    & & & & [0.000]\\
    $\bold{Bitcoin-MSCI\_EM}$ & $\theta=1.443$  & $\theta=1.143$   & $\theta=1.149$   & $\theta=0.233$,\\
    &[0.167]&[0.000]&[0.000]&$\nu=10.499$\\
    & & & &[0.001]\\
    $\bold{Litecoin-MSCI\_W}$ &$\theta=1.352$   & $\theta=1.128$   &  $\theta=1.119$  & $\theta=0.225$,\\
    & [0.088] &[0.000]& [0.009] &$\nu=23.885$\\
    & & & &[0.001]\\
    $\bold{Litecoin-MSCI\_EM}$  &  $\theta=1.485$ & $\theta=1.146$   & $\theta=1.160$ & $\theta=0.248$,\\
    &[ 0.009] & [0.000]& [0.000]&$\nu=26.224$\\
    & & & & [0.000]\\
    \bottomrule
  \end{tabular}
         \begin{tablenotes}
      \item Note: The numbers in brackets represent the $p$-values associated with the test statistics.
    \end{tablenotes}
    \label{tab:8}
\end{table*}

\subsection{\textnormal{\textit{Risk measures results}}\label{4.8}}

 Future risk forecasting is pivotal for risk analysts and policymakers in financial companies and corporations. VaR, ES, and RVaR assessment measures are employed in this research to quantify portfolio risks. This study compares the efficacy of two conventional methodologies (historical simulation and a parametric approach based on the $t$-distribution) with a hybrid GARCH-based copula approach. For performance comparisons, we calculated the averages of the scoring functions for each risk measure. The following steps are adopted in the procedure: (i) the rolling window forecasting technique was used to predict the mean and standard deviation; (ii) the estimation was conducted across the entire sample of asset pairs; (iii) a new period was incorporated into the forecast at the final stage, updating parameters by re-estimating the best-fitted ARMA-eGARCH model for every 10 working days. This approach allows for converting mean and standard deviation to log returns, thereby enabling the anticipation of future performance using the most updated data and periods.
 \par In view of the above, we construct the portfolio by systematically evaluating each asset series and allocating optimal weights described in the principles of modern portfolio theory, assuring an effective approach to achieving risk-adjusted returns\citep{markovitz1959portfolio}. A simulated sample of $10,000$ observations is generated to improve the accuracy of the outcomes produced by the conventional methods under consideration. We generate a dataset comprising $10,000$ observations using estimated copula parameters specific to each copula type for each asset variable. Henceforward, we converted uniformly generated data into standardized residual sequences using induced marginal distributions. The average risk measures for each asset pair with likelihoods of occurrence at 1\%, 2.5\%, and 5\% are presented Table \ref{tab:9}. The findings reveal that the VaR estimates derived from the copula-based EVT model are slightly lower than those obtained from the other two conventional methods. These results imply that the copula-based EVT model performs better than the other two traditional approaches. In Table \ref{tab:9}, we observe that the average RVaR falls between the average VaR values for $\alpha$ and $\beta$, demonstrating that  $\text{VaR}^{\beta}\leq\text{RVaR}^{\alpha, \beta}\leq\text{VaR}^{\alpha}$. Consequently, $\text{RVaR}^{\alpha, \beta}$ is more conservative measure than the $\text{VaR}^{\alpha}$ and suggesting more significant protection than $\text{VaR}^{\beta}$. For example, the values calculated through the frank copula function, we have $\text{VaR}^{5\%}=0.02003<\text{RVaR}^{2.5\%,5\%}=0.02296<\text{VaR}^{2.5\%}=0.02681$. We also realized that $\text{RVaR}^{1\%,5\%}\leq\text{ES}^{\beta}\leq\text{ES}^{\alpha}$ retains for all the computed outcomes. Eventually, we have $\text{VaR}^{\beta}\leq\text{RVaR}^{1\%,5\%}\leq\text{ES}^{\beta}$. Furthermore, as it is compatible with our expectations, we substantiate that the VaR results are lower than the ES results at the same significance levels. Besides, the outcomes acquired from our multivariate risk forecast also adhere to the same principles $\text{RVaR}^{1\%,2.5\%}\geq\text{RVaR}^{1\%,5\%}\geq\text{RVaR}^{2.5\%,5\%}$. As an illustration, the values calculated by the Gumbel copula are 0.03059, 0.02548, and 0.02242 for $\text{RVaR}^{1\%,2.5\%}$, $\text{RVaR}^{1\%,5\%}$, and $\text{RVaR}^{2.5\%,5\%}$ respectively. We observed that lower significance level pairs lead to higher RVaR values. Thus, the highest risk is identified at $\alpha=1\%$ and $\beta=2.5\%$, whereas the lowest forecast is observed at $\alpha=2.5\%$ and $\beta=5\%$. These observances align with the VaR and ES outcomes, indicating that lower significance levels lead to inflated risk projection and enhanced preservation.

\begin{landscape}
\begin{table*}
    \centering
    \caption{Average Portfolio Risk Measures at Significance Levels of 1\%, 2.5\%, and 5\%: VaR, ES, and RVaR.}
    \begin{tabular}{p{4cm}p{2.5cm} p{1.7cm}p{1.7cm}p{1.7cm}p{1.7cm}p{2cm}p{1.7cm}}
        \toprule
        &Risk Measure & Frank & Gumbel & Joe & Student's $t$ & $t$-Distributed & HS \\
        \midrule
Bitcoin-MSCI\_W          & $\text{VaR}^{1\%}$   & 0.03755&0.03684 &0.03622 &0.03906 &0.03971 &0.04674 \\
                         & $\text{VaR}^{2.5\%}$ &0.02681 &0.02623 &0.02565 &0.02725 &0.03332 &0.03469 \\
                         & $\text{VaR}^{5\%}$  &0.02003 &0.01950 &0.01900 &0.01986 &0.02784 &0.02541 \\
                         & $\text{ES}^{1\%}$   &0.05477 &0.05431 &0.05345 &0.05859 &0.06029 &0.07485 \\
                         & $\text{ES}^{2.5\%}$ &0.04063 &0.04008 &0.03937 &0.04271 &0.05298 &0.05474 \\
                         & $\text{ES}^{5\%}$  &0.03179  &0.03124 &0.03062 &0.03288 &0.04582 &0.04180 \\
                         & $\text{RVaR}^{1\%,2.5\%}$ &0.03121 &0.03059 &0.02998&0.03212 &0.03626 &0.04036 \\
                         & $\text{RVaR}^{1\%,5\%}$ &0.02605 &0.02548 &0.02645 &0.02645 &0.03247 &0.03324 \\
                         & $\text{RVaR}^{2.5\%,5\%}$ &0.02296 &0.02242 &0.02305 &0.02305 &0.02991 &0.02892 \\
        \cmidrule{2-8}
Bitcoin-MSCI\_EM         & $\text{VaR}^{1\%}$ &0.04044 &0.03963 &0.03858 &0.03793 &0.04309                          &0.05052 \\
                         & $\text{VaR}^{2.5\%}$ &0.03001 &0.02924 &0.02841 &0.02775 &0.03620 &0.03750 \\
                         & $\text{VaR}^{5\%}$ &0.02325 &0.02252 &0.02183 &0.02125 &0.03031 &0.02659 \\
                         & $\text{ES}^{1\%}$ &0.05708 &0.05654 &0.05520 &0.03252 &0.06568 &0.08028 \\
                         & $\text{ES}^{2.5\%}$ &0.04340 &0.04273 &0.04163 &0.04095 &0.05708 &0.05795 \\
                         & $\text{ES}^{5\%}$ &0.03480 &0.03410 &0.03317 &0.03252 &0.04904 &0.04447 \\
                         & $\text{RVaR}^{1\%,2.5\%}$ &0.03428 &0.03352 &0.03259 &0.03195 &0.04057 &0.04330 \\
                         & $\text{RVaR}^{1\%,5\%}$ &0.02923 &0.02848 &0.02766 &0.02703 &0.03599 &0.03324 \\
                         & $\text{RVaR}^{2.5\%,5\%}$ &0.02619 &0.02546 &0.02470 &0.02408 &0.03292 &0.03106 \\
            \cmidrule{2-8}

Litecoin-MSCI\_W         & $\text{VaR}^{1\%}$   &0.04649 &0.04573 &0.04519 &0.04804 &0.04907 &0.05446 \\
                         & $\text{VaR}^{2.5\%}$ &0.03436 &0.03373 &0.03319 &0.03484 &0.04127 &0.04266 \\
                         & $\text{VaR}^{5\%}$  &0.02586 &0.02530 &0.02476 &0.02571 &0.03459 &0.03350 \\
                         & $\text{ES}^{1\%}$   &0.06001 &0.05949 &0.05858 &0.06365 &0.07524 &0.09032 \\
                         & $\text{ES}^{2.5\%}$ &0.04769 &0.04709 &0.04639 &0.04969 &0.06154 &0.06478 \\
                         & $\text{ES}^{5\%}$  &0.03864  &0.03805 &0.03741 &0.03969 &0.05218 &0.05143 \\
                         & $\text{RVaR}^{1\%,2.5\%}$ &0.03948 &0.03882 &0.03825&0.04039 &0.04454 &0.04812 \\
                         & $\text{RVaR}^{1\%,5\%}$ &0.03329 &0.03269 &0.03212 &0.03370 &0.04059 &0.04193 \\
                         & $\text{RVaR}^{2.5\%,5\%}$ &0.02958 &0.02901 &0.02844 &0.02969 &0.03787&0.03816 \\
        \cmidrule{2-8}
Litecoin-MSCI\_EM        & $\text{VaR}^{1\%}$ &0.04912 &0.04825 &0.04703 &0.05165 &0.05249 &0.05939 \\
                         & $\text{VaR}^{2.5\%}$ &0.03718 &0.03631 &0.03531 &0.03810 &0.04419 &0.04462 \\
                         & $\text{VaR}^{5\%}$ &0.02873 &0.02791 &0.02709 &0.02863 &0.03708 &0.03527 \\
                         & $\text{ES}^{1\%}$ &0.06247 &0.06189 &0.06028 &0.06791 &0.08122 &0.09402 \\
                         & $\text{ES}^{2.5\%}$ &0.05031 &0.04957 &0.04828 &0.05347 &0.06661 &0.06766 \\
                         & $\text{ES}^{5\%}$ &0.04138 &0.04059 &0.03948 &0.04312 &0.05570 &0.05367 \\
                         & $\text{RVaR}^{1\%,2.5\%}$ &0.04220 &0.04135 &0.04027 &0.04384 &0.04754 &0.05045 \\
                         & $\text{RVaR}^{1\%,5\%}$ &0.03611 &0.03527 &0.03429 &0.03692 &0.04314 &0.04378 \\
                         & $\text{RVaR}^{2.5\%,5\%}$ &0.03245 &0.03162 &0.03070 &0.03278 &0.04046 &0.03972 \\
        \bottomrule
\end{tabular}
\label{tab:9}
\end{table*}
\end{landscape}
 \subsection{\textnormal{\textit{Estimation of spillover effects}}\label{4.9}}
In this study, we examine how traditional and digital assets influence forecast error variance by using the strategy described by \citep{diebold2012better}.Table \ref{tab:10} presents the impact of the ten-step-ahead variance decomposition with a VAR model of order one for the entire return series of four assets.
 \begin{table*}
  \centering
  \caption{Results including whole sample size of spillovers }
  \begin{tabular}{p{3.2cm}p{2.1cm}p{2cm} p{2cm} p{2.2cm}p{2cm}}
    \toprule
           &Bitcoin & Litecoin & MSCI\_W & MSCI\_EM& From others except own \\
        \cmidrule{1-6}
    Bitcoin             & 53.96  & 33.55   & 6.68   &  5.81   & 46.04 \\
    Litecoin            & 34.17  & 54.66  & 5.85  &  5.32   & 45.34  \\
    MSCI\_W             & 7.68   & 6.11   & 54.09    & 32.12   & 45.91 \\
    MSCI\_EM            &6.90   & 5.86  &32.61  & 54.63  & 45.37 \\
    To others           & 48.75  & 45.52 & 45.15   & 43.25  &$\bold{182.67}$\\
    To others except own& 102.70   & 100.18  & 99.24 & 97.87  & \\
    & & & & & Spillover Index \\
    & & & & & 45.67\% \\
    \bottomrule
  \end{tabular}
    \label{tab:10}
\end{table*}

The total spillover index in Table \ref{tab:10} is 45.67\%, indicating that 45.67\% of the total forecast error variance in prices arises from spillovers. The off-diagonal components between cryptocurrencies and other financial indices, such as MSCI\_W and MSCI\_EM, are relatively low compared to self-spillovers. For instance, the spillover from Bitcoin to MSCI\_W  is 6.68\%, while the spillover from MSCI\_W to Bitcoin is 7.68\%. These values demonstrate interconnections between markets, though they are not exceptionally strong. The second-to-last row of Table \ref{tab:10} suggests that cryptocurrencies contribute 48.75\% and 45.52\% to the broader global markets, yet they receive less than 50\% from other markets.

\begin{figure*}
  \centering
  \includegraphics[width=162mm, height=115mm]{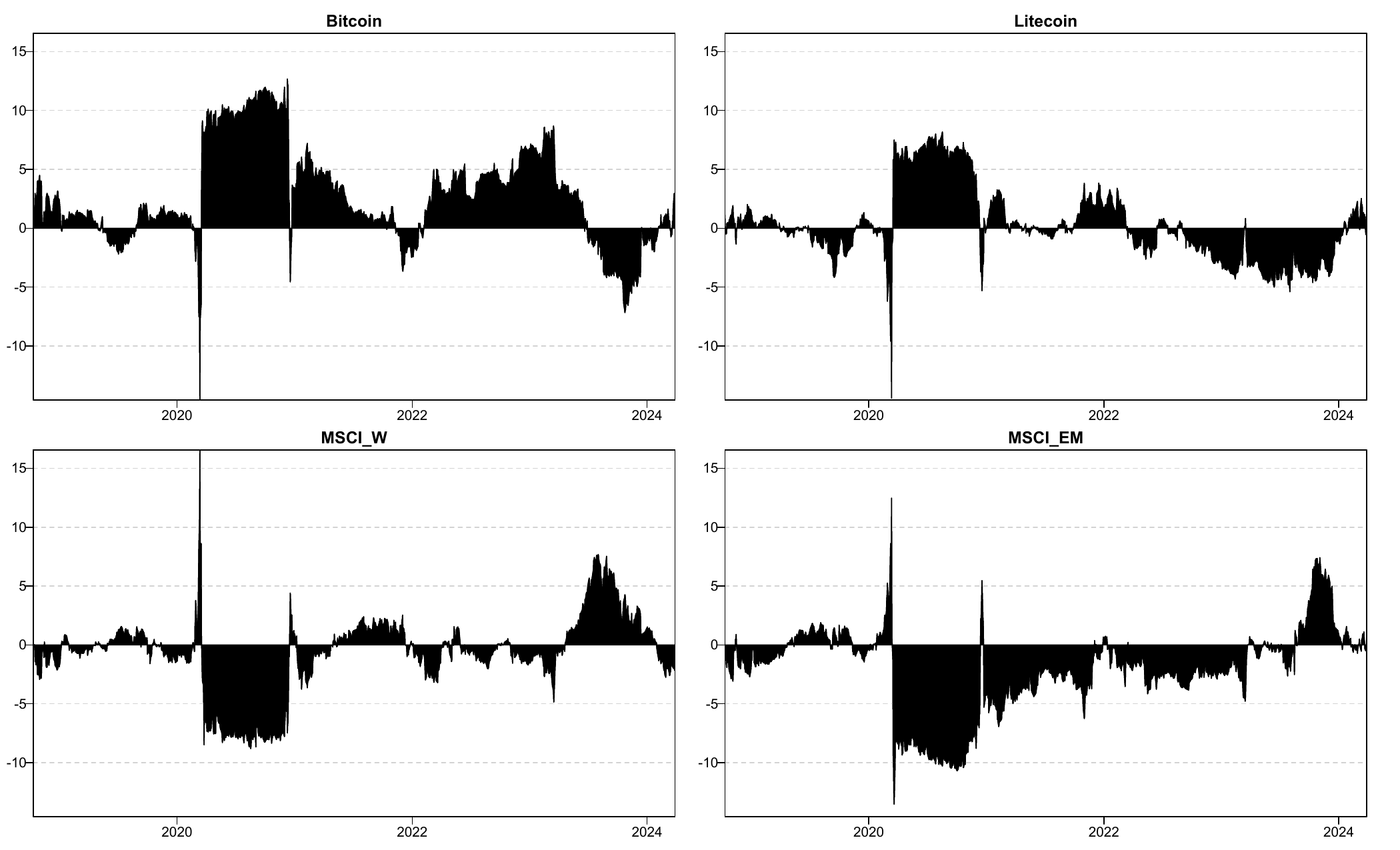} 
\caption{ The impacts of net-spillovers for four asset classes} 
\label{fig:5}
\end{figure*}

Figure~\ref{fig:5} presents the net spillover elements over time for all considered asset returns. The net spillover effect measures the difference between an entity’s influence on other entities and the influence other entities exert on it. This concept is crucial for understanding the dynamics of market influence. Positive values indicate net outgoing spillovers, whereas negative values indicate net incoming spillovers, implying that other commodities more influence a commodity than it influences them. According to Figure~\ref{fig:5}, traditional financial indices exhibit the highest negative values, suggesting substantial external influences. Conversely, cryptocurrencies exhibit high positive values, indicating a significant impact on the broader financial market.

Overall, the findings indicate that cryptocurrencies offer both diversification benefits and challenges in managing new sources of risk within a portfolio containing traditional financial indices. However, they do not provide perfect hedging, particularly during periods of market stress when dependencies are high. This divergence underscores the potential advantages of heterogeneity and the necessity of understanding the fundamental mechanisms generating these spillover effects, which may offer critical insights for portfolio management. The results also argue previous research \citep{rao2022revisiting}, indicating that cryptocurrencies lack robustness as hedging instruments for traditional financial indices, offering diversification potential and mitigating overall risk at a specific expected return level.

\subsection{\textnormal{\textit{Loss function results}}\label{4.10}}
An overview of incurred losses across all combinations of risk measures assumed for individual portfolios can be found in Table \ref{tab:11}. The hybrid \texttt{eGARCH} EVT-based copula functions outperform VaR estimation, while the HS method outperforms ES estimation; accordingly, the $t$-distributed-based parametric approach illustrates superior RVaR computation based on average realized losses. The \texttt{eGARCH} EVT-based copula method reveals the highest realized losses for RVaR despite affecting optimal outcomes for VaR computation across all portfolios. The $t$-distributed-based parametric and HS approaches yield the most elevated realized loss values for VaR and ES. Accordingly, established on these discoveries, it can be intimated that both methods are inappropriate for influential risk management through ES and VaR for portfolios consisting of cryptocurrencies and global financial indices. Among all the implemented copula functions, Joe and Student's $t$ copula functions perform optimally in predicting risk through risk measures across all portfolios. Comparable outcomes were previously documented by \citep{karimi2023analyzing}, who assessed VaR using backtesting procedures and explored cryptocurrencies while substituting global financial indices with commodities like gold and oil prices. 

\par Table \ref{tab:12} presents a comprehensive summary of all the methods contemplated for assessing the risk measures for each portfolio between cryptocurrencies and global financial indices. We further explain these findings in the subsequent ways. First, an essential way that RVaR, VaR, and ES differ is that RVaR is a trimmed risk measure. Because RVaR does not incorporate the most extreme values in its estimation, unlike ES, it furnishes an exceptional perspective in risk assessment. Secondly, incorporating extreme value theory and considering structural dependencies among time series asset pairs, including cryptocurrencies and global financial indices, enhances risk identification, as evidenced by the findings in Table \ref{tab:9}.
\subsection{\textnormal{\textit{Model misspecification and legal robustness analysis}}\label{4.11}}
\par In order to check model misspecification and robustness, Table \ref{tab:13} presents the average measure used to evaluate how different risk measures are analyzed in the context of model misspecification and regulatory robustness. Large discrepancies in risk predictions across various forecast models for the same measure result in significant differences in regulatory capital, which can lead to opportunities for regulatory arbitrage among banks. Our findings indicate that RVaR and VaR exhibit similar changes across different models, suggesting comparable opportunities for regulatory arbitrage. However, ES shows a higher potential for regulatory arbitrage compared to VaR and RVaR. The subsequent investigation by \citet{kellner2016quantifying, muller2022comparison} sheds light on these variations, attributing them to the heightened sensitivity of ES relative to other risk measures. While we include losses above the threshold for RVaR computation, we do not account for losses related to the difficulties in assessing ES. These results indicate that RVaR is a feasible option for forecasting portfolio return risks, potentially improving regulatory stability within banking institutions.

\begin{landscape} 
\begin{table*}
    \centering
    \caption{Average realized loss results for each portfolio under VaR, ES, and RVaR measures at significance levels of 1\%, 2.5\%, and 5\%}
    \begin{tabular}{p{4cm}p{2.5cm} p{1.7cm}p{1.7cm}p{1.7cm}p{1.7cm}p{2cm}p{1.7cm}}
        \toprule
        &Risk Measure & Frank & Gumbel & Joe & Student's $t$ & $t$-Distributed & HS \\
        \midrule
Bitcoin-MSCI\_W          & $\text{VaR}^{1\%}$   &0.04746 &0.04676 &$\textbf{0.04614}$ &0.04898 &0.04962 &0.05665 \\
                         & $\text{VaR}^{2.5\%}$ &0.05171 &0.05114 &$\textbf{0.05056}$ &0.05216 &0.05822 &0.05959 \\
                         & $\text{VaR}^{5\%}$  & 0.06991 &0.06938 &$\textbf{0.06888}$ &0.06975 &0.07771 &0.07529 \\
                         & $\text{ES}^{1\%}$   &0.94602 &0.94644 &0.94731 &0.94221 &$\textbf{0.94048}$ &0.92604 \\
                         & $\text{ES}^{2.5\%}$ &0.96003 &0.96059 &0.96127 &0.95792 &0.94772 &$\textbf{0.94602}$ \\
                         & $\text{ES}^{5\%}$  &0.96893  &0.96951 &0.97012 &0.96778 &$\textbf{0.95487}$ &0.95884 \\
                         & $\text{RVaR}^{1\%,2.5\%}$ &1.01022 & 1.01023 &1.01021&1.01016 &1.01005 &$\textbf{1.01004}$ \\
                         & $\text{RVaR}^{1\%,5\%}$ &1.01069 &1.01070 &1.01069 &1.01154 &$\textbf{1.01004}$ &1.01024 \\
                         & $\text{RVaR}^{2.5\%,5\%}$ &1.02596 &1.02597 &1.02596 &1.02630 &$\textbf{1.02531}$ &1.02552 \\
        \cmidrule{2-8}
Bitcoin-MSCI\_EM         & $\text{VaR}^{1\%}$ &0.05038 &0.04956 &0.04851 &$\textbf{0.04787}$ &0.05302 &0.06044 \\
                         & $\text{VaR}^{2.5\%}$ &0.05493 &0.05417 &0.05334 &$\textbf{0.05269}$ &0.06113 &0.06242 \\
                         & $\text{VaR}^{5\%}$ & 0.07316 &0.07243 &0.07174 &$\textbf{0.07116}$ &0.08021 &0.07529 \\
                         & $\text{ES}^{1\%}$ &0.94386 &0.94435 &0.94565 &0.94637 &0.93518 & $\textbf{0.92075}$ \\
                         & $\text{ES}^{2.5\%}$ &0.95738 &0.95801 &0.95910 &0.95975 &0.94373 &$\textbf{0.94297}$ \\
                         & $\text{ES}^{5\%}$ &0.96620 &0.96691 &0.96783 &0.96851 &$\textbf{0.95180}$ &0.95617 \\
                         & $\text{RVaR}^{1\%,2.5\%}$ &1.01013 &1.01014 & 1.01023 &1.01024 &1.01005 &$\textbf{1.01004}$ \\
                         & $\text{RVaR}^{1\%,5\%}$ &1.01070 &1.01071 &1.01087 &1.01087 &$\textbf{1.01003}$ &1.01015 \\
                         & $\text{RVaR}^{2.5\%,5\%}$ &1.02598 &1.02597 &1.02614 &1.02615 &$\textbf{1.02531}$ &1.02541 \\

        \cmidrule{2-8}
Litecoin-MSCI\_W         & $\text{VaR}^{1\%}$ &0.05646 &0.05570 &$\textbf{0.05515}$ &0.05801 &0.05903 &0.06442 \\
                         & $\text{VaR}^{2.5\%}$ &0.05927 &0.05864 &$\textbf{0.05810}$&0.05974 & 0.06622 &0.05959 \\
                         & $\text{VaR}^{5\%}$ &0.07581 &0.07525 &$\textbf{0.07471}$ &0.07566 &0.08453 &0.08344 \\
                         & $\text{ES}^{1\%}$ &0.94132 &0.94179 &0.94272 &0.93772 &0.92584 &$\textbf{0.91072}$ \\
                         & $\text{ES}^{2.5\%}$ &0.95388 &0.95447 &0.95517 & 0.95177 &0.93969 &$\textbf{0.93646}$ \\
                         & $\text{ES}^{5\%}$ &0.96327 &0.96386 &0.96449 &0.96213 &$\textbf{0.94940}$ &0.95011 \\
                         & $\text{RVaR}^{1\%,2.5\%}$ &1.01096 &1.01096 &1.01096 &1.01084 &1.01005 &$\textbf{1.01004}$\\
                         & $\text{RVaR}^{1\%,5\%}$ &1.01128 &1.01140 &1.01140 &1.01140 &1.01082 &$\textbf{1.01024}$ \\
                         & $\text{RVaR}^{2.5\%,5\%}$ &1.02655 &1.02667 &1.02666 &1.02655 &1.02610 &$\textbf{1.02609}$ \\
        \cmidrule{2-8}
Litecoin-MSCI\_EM        & $\text{VaR}^{1\%}$ & 0.05911 &0.05824 &$\textbf{0.05702}$ &0.06164 &0.06247 &0.06938 \\
                         & $\text{VaR}^{2.5\%}$ &0.06217 &0.06130 &0.06032 &$\textbf{0.06030}$ &0.06917 &0.06242 \\
                         & $\text{VaR}^{5\%}$ &0.07872 &0.07790 &$\textbf{0.07708}$ &0.07862 &0.08706 &0.08525 \\
                         & $\text{ES}^{1\%}$ &0.93906 &0.93961 &0.94112 &0.93367 &0.91997 &$\textbf{0.90739}$ \\
                         & $\text{ES}^{2.5\%}$ &0.95135 &0.95210 &0.95340 &0.94806 &0.93465 &$\textbf{0.93360}$ \\
                         & $\text{ES}^{5\%}$ &0.96074 &0.96156 &0.96267 &0.95890 &$\textbf{0.94591}$ &0.94791 \\
                         & $\text{RVaR}^{1\%,2.5\%}$ &1.01096 &1.01111 &1.01096 &1.01083 &$\textbf{1.01004}$ &1.01005 \\
                         & $\text{RVaR}^{1\%,5\%}$ &1.01150 &1.01149 &1.01151 & 1.01149 &1.01097 &$\textbf{1.01015}$ \\
                         & $\text{RVaR}^{2.5\%,5\%}$ &1.02677 &1.02676 &1.02675 &1.02678 &$\textbf{1.02625}$ &1.02626 \\
        \bottomrule
    \end{tabular}
\label{tab:11}
\end{table*}
\end{landscape}

\begin{table*}
    \centering
    \caption{Overview of optimal models identified by risk measures at significance levels of 1\%, 2.5\%, and 5\%}
    \begin{tabular}{p{2.8cm}p{2.7cm} p{3cm}p{2.9cm}p{2.3cm}}
        \toprule
        Risk Measure & Bitcoin-MSCI\_W & Bitcoin-MSCI\_EM & Litecoin-MSCI\_W & Litecoin-MSCI\_EM\\
        \midrule
 $\text{VaR}^{1\%}$ &Joe &Student's $t$ &Joe &Joe\\
 $\text{VaR}^{2.5\%}$ &Joe &Student's $t$ &Joe&Student's $t$ \\
$\text{VaR}^{5\%}$  &Joe &Student's $t$  &Joe &Joe \\
$\text{ES}^{1\%}$   &$t$-Distributed  &HS &HS &HS  \\
  $\text{ES}^{2.5\%}$ &HS &HS &HS &HS  \\
$\text{ES}^{5\%}$  &$t$-Distributed &$t$-Distributed&$t$-Distributed &$t$-Distributed \\ 
$\text{RVaR}^{1\%,2.5\%}$ &HS &HS  &HS&$t$-Distributed \\
$\text{RVaR}^{1\%,5\%}$ &$t$-Distributed&$t$-Distributed&HS &HS \\
$\text{RVaR}^{2.5\%,5\%}$ &$t$-Distributed &$t$-Distributed &HS &$t$-Distributed \\
        \bottomrule
    \end{tabular}
    \label{tab:12}
\end{table*}

\begin{table*}
    \centering
    \caption{Analysis of legal robustness and model misspecification in risk forecasting: A comparison of VaR, ES, and RVaR across different significance levels}
    \begin{tabular}{p{3.7cm}p{2.7cm} p{3cm}p{2.7cm}p{1.7cm}}
        \toprule
        Risk Measure & Bitcoin-MSCI\_W & Bitcoin-MSCI\_EM & Litecoin-MSCI\_W & Litecoin-MSCI\_EM\\
        \midrule
 $\text{VaR}^{1\%}$         &0.12818  &0.13170  &0.08407  &0.08896\\
 $\text{VaR}^{2.5\%}$       &0.15771  & 0.14747  &0.11235 &0.10124\\
$\text{VaR}^{5\%}$          &0.17799  &0.14387  &0.14734 &0.10130\\
$\text{ES}^{1\%}$           & 0.15603 &0.17575  &0.15858 &0.16350 \\
  $\text{ES}^{2.5\%}$       &0.17468  & 0.18302  &0.14121 & 0.14289\\
$\text{ES}^{5\%}$           &0.19008  & 0.18605 &0.14887  &0.11805\\ 
$\text{RVaR}^{1\%,2.5\%}$   &0.14605  &0.14749  & 0.09751 &0.09129 \\
$\text{RVaR}^{1\%,5\%}$     &0.15544  & 0.13303 & 0.11803 & 0.10412\\
$\text{RVaR}^{2.5\%,5\%}$   &0.16205  & 0.14371 & 0.13454 & 0.09521\\
        \bottomrule
    \end{tabular}
    \label{tab:13}
\end{table*}

\section{Conclusion and policy recommendations}\label{consclusion}
This study aims to quantify risk spillovers across assets and derive risk measures like VaR, ES, and RVaR for portfolios containing cryptocurrencies and global financial indices. We compare conventional methods (HS and $t$-distribution) with a hybrid \texttt{eGARCH} EVT-based copula method. We begin by selecting four sequences of digital and traditional assets, confirming their heavy-tailed, leptokurtic, and non-$i.i.d$ properties. We model these using ARMA-GARCH, finding \texttt{eGARCH} most effective for capturing volatility. For further analysis, standardized residuals are derived for each series to achieve approximate $i.i.d$ properties. In the copula phase, various heavy-tailed probability distributions are evaluated, and the optimal one is selected using statistical criteria. The dependency structure among assets is estimated using Frank, Gumbel, Joe, and Student's $t$ copulas. We identify the best copula function through diverse loss functions, considering model misspecification and legal robustness. Our findings reveal that the \texttt{eGARCH} EVT-based copula method is more optimistic for VaR estimation, while HS and $t$-distribution methods are more suitable for ES and RVaR, respectively. ES shows greater sensitivity to model misspecification compared to VaR and RVaR. RVaR proves to be a significant risk measure for volatile portfolios, aiding regulatory balance among banking institutions.
\\
The study finds minimal risk spillover from cryptocurrencies to global financial assets, challenging their perceived safety and highlighting the exploitation of volatility spillovers in diversified portfolios. The \texttt{eGARCH} EVT-based copula model enhances VaR accuracy, while HS and $t$-distribution methods improve ES and RVaR estimates. Portfolio managers should consider tail risk dependencies when combining digital and traditional assets. Due to the adequate performance of the Bulk and Tails parametric model (i.e., BATs) used in this study, we recommended use for risk management. This entertains the entire data of return residual and models the extreme tail behaviors without considering any threshold for tails. Therefore, the approach proposed in this study is considered superior to models discussed in \citet{karimi2023analyzing}. RVaR, an improved ES measure, offers robust risk assessment, reduces model risk, and provides stability against outliers. RVaR also aids regulators in managing volatility, enhancing financial system resilience, and aligning with VaR by accounting for extreme losses, thereby providing comprehensive risk assessments and supporting a harmonized regulatory environment.
\\
The risk spillover analysis reveals a slight dependency and strong concordance between financial indices and cryptocurrencies, suggesting cryptocurrencies can serve as diversifiers in portfolio optimization. This insight allows risk management authorities to construct portfolios with both traditional and digital assets to manage volatility more effectively. Despite offering diversification benefits, cryptocurrencies, particularly Litecoin due to its correlation with Bitcoin, do not provide superior hedging advantages and cannot withstand market volatility independently. The heightened spillover indices highlight the need to optimize portfolios for both risk and volatility, leveraging the unique properties of cryptocurrencies alongside global financial indices.
\\
While our study significantly contributes to the literature, future research can expand by including additional asset classes such as equities, treasury bills, and commodities-based portfolios to forecast risk measures like VaR, ES, and RVaR and test their robustness. Further investigation can explore integrating volatility models with time-varying copula functions to analyze the long-term memory effect on the dependence structure and estimate risk measures. Additionally, employing novel extreme value probability models within a univariate framework to compute risk metrics at various confidence levels is a promising direction. Future comparative studies should evaluate risk measures based on loss functions and consider the impact of legal robustness across a broader spectrum of models.
\\
In summary, the \texttt{eGARCH} EVT-based copula model provides distinct interpretations of VaR compared to standard methods, while ES demonstrates increased sensitivity to regulatory arbitrage, aligning with VaR and RVaR. These findings have significant implications for investors, portfolio managers, and policymakers. The study's policy recommendations aim to mitigate adverse risk spillover effects between cryptocurrencies and traditional markets, thereby promoting market stability. These benchmarks enhance financial system robustness and ensure smoother market operations.

\printcredits

\section*{Declaration of competing interest}
The authors declare that they have no competing financial interests or personal relationships that could have influenced the content of this study.

\section*{Data and code availability:} Data and code for this study are available upon reasonable request.

\section*{Acknowledgment}  
This research is partially funded by the Technology and Innovation Major Project of the Ministry of Science and Technology of China (Grant No. 2020AAA0108402) and by the National Natural Science Foundation of China (NSFC) under Grants 72210107001 and 71825007. T.~Ahmad acknowledges support from the R\'egion Bretagne through project SAD-2021-MaEVa.

\bibliographystyle{cas-model2-names}
\bibliography{cas-refs_RP}

\end{document}